\documentclass[a4paper,11pt]{article}
\usepackage{jheppub} 

\usepackage{caption}    
\usepackage{float}       
\usepackage{array}
\usepackage{booktabs}
\usepackage{multirow}
\usepackage[table]{xcolor}
\newtheorem{theorem}{Theorem}
\allowdisplaybreaks[1]
\arxivnumber{2507.11315} 

\title{\boldmath Deformed Schur Indices of $BCD$-type   for $\mathcal{N} = 4$ Super Yang-Mills and Symmetric Functions}
\preprint{
    USTC-ICTS/PCFT-25-28
}

\author[a]{Gao-fu Ren} \author[b]{and Min-xin Huang}
\affiliation[a]{Interdisciplinary Center for Theoretical Study, University of Science and Technology of China,  Hefei, Anhui 230026, China}
\affiliation[b]{Peng Huanwu Center for Fundamental Theory,   Hefei, Anhui 230026, China}

\emailAdd{rengaofu@mail.ustc.edu.cn, minxin@ustc.edu.cn}

\abstract{We investigate the deformed Schur index in four dimensional $\mathcal{N}=4$ super Yang-Mills theories with $SO$ and $Sp$ gauge groups, generalizing Hatsuda's recent calculations. We express the deformed Schur index as integrals of Koornwinder polynomials and Macdonald polynomials, then perform the integrals in terms of the normalization constants of Macdonald polynomials. We provide explicit results for some low rank gauge groups and for expansion in a $u$ parameter. We discuss various special limits and the tests of S-duality. }

\begin{document}
\maketitle
\flushbottom
\section{Introduction}

As a type of Witten index \cite{witten:1982df}, superconformal indices are powerful invariants that encode protected spectral data of superconformal field theories \cite{Romelsberger:2005eg, Kinney_2007}. They serve as useful tools for precise tests of field theory dualities, such as Seiberg duality, S-duality, see e.g. the review papers  \cite{Rastelli:2016tbz, Gadde:2020yah}. They have also been instrumental in counting microstates of supersymmetric black hole in $AdS_5$ space through the holographic dual 4d CFT \cite{Cabo-Bizet:2018ehj, Choi:2018hmj, Benini:2018ywd}. These novel developments have entailed the interesting problem of  distinguishing the graviton states and non-graviton states,  or fortuitous states, e.g. discussed in \cite{Chang:2022mjp, Choi:2023znd, Chang:2024zqi}. 

We will consider the  $\mathcal{N}=4$ maximally supersymmetric Yang-Mills (SYM) theory, which is most relevant for holographic black hole microstate counting, as well as tests of the giant graviton expansion proposed in \cite{Arai:2020qaj, Imamura:2021ytr, Gaiotto:2021xce, Murthy:2022ien}. Giant gravitons are D-branes in AdS space described by determinant operators in the dual CFT. They were first studied in the early days of AdS/CFT correspondence in \cite{McGreevy:2000cw, Grisaru:2000zn, Hashimoto:2000zp, Balasubramanian:2001nh, Corley:2001zk}, and have seen some recent revivals in e.g. \cite{Jiang:2023uut, Chen:2025yxg}. The giant graviton expansion interprets the coefficients in the expansion of the superconformal index in a fugacity parameter as contributions of D-branes. However, the quantitative  tests of the relations usually require analytic continuation of the fugacity parameters, so some closed formulas of the superconformal index are urgently needed.

The most general superconformal index in the  $\mathcal{N}=4$  SYM theory has 4 independent fugacity parameters, and can be rewritten as specific elliptic hypergeometric integrals \cite{MR2479997, Spiridonov_2011}. However the integrals are in general difficult to compute. Instead, we consider a 3-parameter deformed Schur index, recently studied by Hatsuda \cite{hatsuda2025deformedschurindicesmacdonald}. The deformed Schur indices can be written as integrals of $q$-Pochhammer symbols and are simpler to compute. In further special limits, the deformed Schur indices reduced to some even simpler cases, such as the 1/4 BPS index, the 1/8 BPS Schur index \cite{Gadde:2011uv, Bourdier:2015wda}, half index \cite{Gaiotto:2019jvo}. In particular, the Schur index has nice modular properties, and has been studied in \cite{Pan:2021mrw, Beem:2021zvt, Huang:2022bry, du2023schurindicesmathcaln4superyangmills}.  The deformed Schur indices and its various limit has also been quite useful to provide tests for the giant graviton expansion, see e.g. \cite{Hatsuda:2022xdv,  Beccaria:2024szi,  Hatsuda:2024uwt, hatsuda2025deformedschurindicesmacdonald}.

In \cite{hatsuda2025deformedschurindicesmacdonald}, using the properties of Macdonald polynomials, Hatsuda obtained some nice formulas for the deformed Schur index, characterized by three parameters $(t,u,q)$, greatly simplify previous results in the literature. Some useful properties of symmetric polynomials can be found in the books \cite{macdonald1998symmetric, noumi2023macdonald}. Although the defining integrals of deformed Schur index is symmetric in the $t$ and $u$ variables, the symmetry is no longer manifest in Hatsuda's formulas but can be explicitly checked, providing some non-trivial identities of $q$-Pochhammer symbols. In this paper, we  investigate the deformed Schur indices for $BCD$-type gauge groups in $\mathcal{N}=4$ SYM, formulated as  integrals of the specializations of Koornwinder polynomials.  The $SO(2N+1)$ case requires particular attention: its exact form derives from augmenting the $SO(2N)$ integral with additional terms and $B_N$ type weight function. Although the calculations are much more complicated than the $SU(N)$ case in \cite{hatsuda2025deformedschurindicesmacdonald}, we also manage to write the deformed Schur index in terms of the normalization factors $\mathcal{N}_{\lambda,N}$ of Macdonald polynomials, where $\lambda$ is an integer partition. We provide explicit results for some low rank gauge groups and for expansion in the $u$ parameter.

The Koornwinder polynomials are the most general orthogonal polynomials associated to the nonreduced $BC_n$ root systems,  and reduce to Macdonald polynomials of types $B_n, C_n, D_n$ in certain special cases. They have also appeared in some previous research on superconformal indices in different physical contexts.  The papers \cite{Mekareeya:2012tn, Lemos:2012ph} considered class $\mathcal{S}$ theories of type $D$ from the 6d $(2,0)$ theories compactified in Riemann surfaces with various twists. They computed the superconformal indices in various special limits using the TQFT (topological quantum field theory) on the Riemann surfaces.  The papers \cite{Nazzal:2018brc, Nazzal:2021tiu} considered  6d $(1,0)$ theories (with conformal matter) compactified in Riemann surfaces. They studied the insertion of surface defects via the van Diejen difference operator, also computed the index in special limits. Here we focus on the more special theory of $\mathcal{N}=4$ super Yang-Mills but the parameter space for deformed Schur index is more general.

This paper is organized as follows. Section \ref{sec2} presents a review of established results concerning the deformed Schur indices in $\mathcal{N}=4$ SYM theories with $BCD_N$-type gauge groups. We further introduce a Koornwinder-type integral that serves as a natural generalization of these deformed Schur indices and establish key combinatorial properties of Koornwinder polynomials that will be essential for our subsequent analysis. In Section \ref{sec3}, we compute the deformed Schur indices using the decomposition of the products of monomial symmetric function and  Macdonald polynomials. The calculations in this section do not use the Koornwinder-type integrals, The results can be expanded in terms of the normalization constants of Macdonald polynomials, but are not convenient for expansion in the fugacity parameters.  In Section \ref{sec4}, we use  the Koornwinder polynomial properties exhibited in Section \ref{sec2} to derive a formal series expansion of the deformed Schur indices in the fugacity parameter $u$. In Section \ref{sec5}, we examine the behavior of the deformed Schur indices under various further limits and provide an explicit test of S-duality. In section \ref{sec6}, we conclude by summarizing our work and proposing several promising directions for future research. In Appendix A, we give some notation in Young tableau and symmetric functions. In Appendix B, we list some results about Macdonald polynomial expansion of symmetric function products which are essential in this paper.

\section{$BCD_N$-type deformed Schur indices of $\mathcal{N}=4$ SYM theories}
\label{sec2}
We start with the elliptic hypergeometric integrals representation of the superconformal indices in $\mathcal{N}=4$ $Sp(2N)$, $SO(2N+1)$, $SO(2N)$ SYM theories \cite{Spiridonov_2011} :
\begin{align}
\label{Sp(2N)SCI}
&I_{Sp(2N)}^{SCI} = \chi'_N \int_{\mathbb{T}^N} 
\prod_{1 \leq i < j \leq N} \frac{\prod_{k=1}^3 \Gamma(s_k z_i^{\pm 1} z_j^{\pm 1}; p, q)}{\Gamma(z_i^{\pm 1} z_j^{\pm 1}; p, q)} 
\prod_{j=1}^N \frac{\prod_{k=1}^3 \Gamma(s_k z_j^{\pm 2}; p, q)}{\Gamma(z_j^{\pm 2}; p, q)} 
\frac{dz_j}{2\pi i z_j},\\[2ex]
\label{SO(2N+1)SCI}
&I_{SO(2N+1)}^{SCI} = \chi'_N \int_{\mathbb{T}^N} 
\prod_{1 \leq i < j \leq N} \frac{\prod_{k=1}^3 \Gamma(s_k z_i^{\pm 1} z_j^{\pm 1}; p, q)}{\Gamma(z_i^{\pm 1} z_j^{\pm 1}; p, q)} 
\prod_{j=1}^N \frac{\prod_{k=1}^3 \Gamma(s_k z_j^{\pm 1}; p, q)}{\Gamma(z_j^{\pm 1}; p, q)} 
\frac{dz_j}{2\pi i z_j},\\[2ex]
\label{SO(2N)SCI}
&I_{SO(2N)}^{SCI} = \chi'_N \int_{\mathbb{T}^N} 
\prod_{1 \leq i < j \leq N} \frac{\prod_{k=1}^3 \Gamma(s_k z_i^{\pm 1} z_j^{\pm 1}; p, q)}{\Gamma(z_i^{\pm 1} z_j^{\pm 1}; p, q)} 
\prod_{j=1}^N \frac{dz_j}{2\pi i z_j},
\end{align}
where
\begin{align*}
&\Gamma(a, b; p, q) := \Gamma(a; p, q)\Gamma(b; p, q), \quad\Gamma(az^{\pm 1}; p, q) := \Gamma(az; p, q)\Gamma(az^{-1}; p, q),\\
&\Gamma(z_i^{\pm}z_j^{\pm}; p, q) := \Gamma(z_i^{+1}z_j^{+1}; p, q)\Gamma(z_i^{+1}z_j^{-1}; p, q)\Gamma(z_i^{-1}z_j^{+1}; p, q)\Gamma(z_i^{-1}z_j^{-1}; p, q),
\end{align*}
and
\begin{equation*}
\chi'_N  = \left\{
    \begin{aligned}
    & \frac{(p;p)_\infty^N (q;q)_\infty^N}{2^{N}N!} \prod_{k=1}^3 \Gamma^N(s_k; p, q) \ && \text{$Sp(2N)$ or $SO(2N+1)$}, \\
    & \frac{(p; p)_\infty^N (q; q)_\infty^N}{2^{N-1} N!} \prod_{k=1}^3 \Gamma^N(s_k; p, q) && \text{$SO(2N)$}. \\
    \end{aligned}
\right.
\end{equation*}
In our notation, we take $s_1=t$, $s_2=u$, $s_3=v$, and these five parameters are not independent, there is a constraint $tuv=pq$, to get integral representations for the deformed Schur indices, with the properties of the elliptic gamma function:
\begin{align*}
    \Gamma(a, b; p, q)=1\ \ \text{(if $ab=pq$)},\quad\quad\Gamma(z; p, q) \underset{p\to 0}{=} \frac{1}{(z; q)_\infty}  \underset{q\to 0}{=} \frac{1}{1-z},
\end{align*}
we get the deformed Schur indices of $\mathcal{N}=4$ $Sp(2N)$, $SO(2N+1)$, $SO(2N)$ SYM theories as the integrals of $q$-Pochhammer symbol by taking the limit $p=v=0$ :
\begin{align}\label{Cdsi}
I_{Sp(2N)} (u,q,t)  &=
\begin{aligned}[t]
    &\frac{(q; q)_{\infty}^N}{2^N N!} \frac{(t u; q)_{\infty}^N}{(t, u; q)_{\infty}^N}\\
&\times \oint_{\mathbb{T}^N} \prod_{1 \leq i < j \leq N} \frac{(z_i^{\pm 1} z_j^{\pm 1}, t u z_i^{\pm 1} z_j^{\pm 1}; q)_{\infty}}{(t z_i^{\pm 1} z_j^{\pm 1}, u z_i^{\pm 1} z_j^{\pm 1}; q)_{\infty}} \prod_{j=1}^N \frac{(z_j^{\pm 2}, t u z_j^{\pm 2}; q)_{\infty}}{(t z_j^{\pm 2}, u z_j^{\pm 2}; q)_{\infty}} \frac{dz_j}{2\pi iz_j},
\end{aligned}\\[2ex]
\label{Bdsi}
I_{SO(2N+1)} (u,q,t)  &= 
\begin{aligned}[t]
&\frac{(q; q)_{\infty}^N}{2^N N!} \frac{(t u; q)_{\infty}^N}{(t, u; q)_{\infty}^N}\\
&\times \oint_{\mathbb{T}^N} \prod_{1 \leq i < j \leq N} \frac{(z_i^{\pm 1} z_j^{\pm 1}, t u z_i^{\pm 1} z_j^{\pm 1}; q)_{\infty}}{(t z_i^{\pm 1} z_j^{\pm 1}, u z_i^{\pm 1} z_j^{\pm 1}; q)_{\infty}} \prod_{j=1}^N \frac{(z_j^{\pm 1}, t u z_j^{\pm 1}; q)_{\infty}}{(t z_j^{\pm 1}, u z_j^{\pm 1}; q)_{\infty}} \frac{dz_j}{2\pi iz_j},
\end{aligned}\\[2ex]
 \label{Ddsi}
I_{SO(2N)} (u,q,t)  &=\frac{(q; q)_{\infty}^N}{2^{N-1} N!} \frac{(t u; q)_{\infty}^N}{(t, u; q)_{\infty}^N}
\times \oint_{\mathbb{T}^N}\prod_{1 \leq i < j \leq N} \frac{(z_i^{\pm 1} z_j^{\pm 1}, t u z_i^{\pm 1} z_j^{\pm 1}; q)_{\infty}}{(t z_i^{\pm 1} z_j^{\pm 1}, u z_i^{\pm 1} z_j^{\pm 1}; q)_{\infty}}  \prod_{j=1}^N\frac{dz_j}{2\pi iz_j}.
\end{align}

As mentioned in \cite{hatsuda2025deformedschurindicesmacdonald}, the deformed Schur index for $U(N)$ $\mathcal{N}=4$ SYM theory is:
\begin{equation*}
    I_{U(N)}(t,u;q) = \frac{1}{N!} \frac{(q;q)_\infty^N (tu;q)_\infty^N}{(t;q)_\infty^N (u;q)_\infty^N} \oint_{\mathbb{T}^N} \prod_{i=1}^N \frac{dx_i}{2\pi i x_i} 
\prod_{1 \leq i \neq j \leq N} \frac{(x_i/x_j;q)_\infty (tu x_i/x_j;q)_\infty}{(t x_i/x_j;q)_\infty (u x_i/x_j;q)_\infty}.  
\end{equation*}
The integrand can be naturally decomposed into two distinct components, one is the weight function: 
\begin{equation}
\label{weightfunction}
    w(\mathbf{x}) = \prod_{1 \leq i \neq j \leq N} \frac{(x_i/x_j;q)_\infty }{(t x_i/x_j;q)_\infty },
\end{equation}
and the remaining part associated with Macdonald polynomials:
\begin{equation}
\label{macintegrand}
  \prod_{i,j=1}^N \frac{(tu x_i/x_j;q)_\infty}{(u x_i/x_j;q)_\infty} 
= \sum_{\ell(\lambda) \leq N} u^{|\lambda|} b_\lambda P_\lambda(\mathbf{x};q,t) P_\lambda(\mathbf{x}^{-1};q,t).
\end{equation}
The integral with weight function (\ref{weightfunction}) define a scalar product of symmetric function \cite{noumi2023macdonald}. Then the  Macdonald polynomials are orthogonal with respect to this scalar product:
\begin{equation}
\label{scpro}
\frac{1}{N!} \oint_{\mathbb{T}^N} \prod_{i=1}^N \frac{dx_i}{2\pi i x_i} 
w(\mathbf{x}) P_\lambda(\mathbf{x};q,t) P_\mu(\mathbf{x}^{-1};q,t) 
= \mathcal{N}_{\lambda,N}\delta_{\lambda \mu},
\end{equation}
where the normalization constant $\mathcal{N}_{\lambda,N}$ is :
\begin{equation}
\label{normfactor}
\mathcal{N}_{\lambda,N} = \prod_{1\leq i<j\leq N} 
\frac{(t^{j-i}q^{\lambda_i-\lambda_j+1};q)_\infty
(t^{j-i}q^{\lambda_i-\lambda_j};q)_\infty}
{(t^{j-i+1}q^{\lambda_i-\lambda_j};q)_\infty
(t^{j-i-1}q^{\lambda_i-\lambda_j+1};q)_\infty}.
\end{equation}
Using these formulas we can find the deformed Schur index for $U(N)$ is:
\begin{equation*}
I_N(t,u;q) = \frac{(q;q)_\infty^N}{(t;q)_\infty^N} 
\sum_{\ell(\lambda) \leq N} u^{|\lambda|} b_\lambda \mathcal{N}_{\lambda,N}.
\end{equation*}
Analogously,  the deformed Schur indices admit a reformulation consisting of two parts: a weight function part and remaining part involving $CD_N$ type Macdonald polynomials for $CD_N$ type deformed Schur indices,  the $B_N$ case exhibits subtle distinctions in its algebraic structure, but all cases ultimately reduce to combinatorial expressions involving Macdonald polynomial. More generally, we can introduce an integral  using Koornwinder polynomials. 
Let \(\Delta_n(\mathbf{x};q,t;t_0,t_1,t_2,t_3) \) denote the Koornwinder weight function:
\[\Delta_n(x_1,x_2,\ldots,x_n;q,t;t_0,t_1,t_2,t_3) = 
\prod_{1 \leq i \leq n} \frac{(x_i^{2};q)_\infty}
{(t_0 x_i, t_1 x_i, t_2 x_i, t_3 x_i;q)_\infty} 
\prod_{1 \leq i < j \leq n} \frac{(x_i x_j^\pm;q)_\infty}
{(t x_i x_j^\pm;q)_\infty}.
\]
To reformulate the deformed Schur indices, we need some theorems about Koornwinder polynomials\cite{MR2127341}\cite{MR4259867}:
\begin{theorem}\label{thm1}
For $m$ a non-negative integer or half-integer and $\lambda$ a partition,
\begin{align*}
\lim_{m \to \infty} (x_1 \cdots x_n)^m K_{m^n - \lambda}&(\mathbf{x}; q, t; t_0, t_1, t_2, t_3)
=\\
&P_\lambda(\mathbf{x}; q, t) \prod_{i=1}^n \frac{(t_0 x_i, t_1 x_i,t_2 x_i, t_3 x_i; q)_\infty}{(x^2_i; q)_{\infty}}
\prod_{1 \leq i < j \leq n} \frac{(t x_i x_j; q)_\infty}{(x_i x_j; q)_\infty}.
\end{align*}
\end{theorem}
\noindent Consider the integral:
\begin{align}
\label{KDSI}
\begin{split}
\ I'_n(u,q,t&;t_0,t_1,t_2,t_3) = \\
&\mathcal{X}_n \lim_{m \to \infty}\sum_{\lambda}b_{\lambda}(q,t)\oint_{\mathbb{T}^n}u^{mn} \ \Delta_n(\mathbf{x};q,t;t_0,t_1,t_2,t_3) \Delta_n(\mathbf{x^{-1}};q,t;t_0,t_1,t_2,t_3) \\
&\times K_{m^n - \lambda}(\sqrt{u}\mathbf{x}; q, t; t_0,t_1,t_2,t_3) K_{m^n - \lambda}(\sqrt{u}\mathbf{x}^{-1}; q, t; t_0,t_1,t_2,t_3)\prod_{j=1}^N\frac{dx_j}{2\pi ix_j},
\end{split}
\end{align}
all objects in this integral is invariant under the permutations of $(t_0,t_1,t_2,t_3)$.
 And in special chosen of $(t_0,t_1,t_2,t_3)$ and $\mathcal{X}_n$: 
\begin{align*}
(t_0,t_1,t_2,t_3)&= (+\sqrt{t},-\sqrt{t},+\sqrt{qt},-\sqrt{qt}),\quad
&\mathcal{X}_n &= \frac{(q; q)_{\infty}^n}{2^n n!(t; q)_{\infty}^n} &for\ \ C_n,\\
(t_0,t_1,t_2,t_3)&= (-1,+t,-\sqrt{q},+\sqrt{q}),&\mathcal{X}_n &= \frac{(q; q)_{\infty}^n}{2^n n!(t; q)_{\infty}^n} &for\ \ B_n,\\
(t_0,t_1,t_2,t_3)&= (-1,+1,-\sqrt{q},+\sqrt{q}),&\mathcal{X}_n &= \frac{(q; q)_{\infty}^n}{2^{n-1} n!(t; q)_{\infty}^n}  &for\ \ D_n,
\end{align*}
we can find that $I'_{C_N}=I_{Sp(2N)}$ and $I'_{D_N}=I_{SO(2N)}$, although $I_{SO(2N+1)}\neq I'_{B_N}$, it can be reformulated  through the introduction of appropriate $q$-Pochhammer symbols in the part involving $D_N$ type Koornwinder polynomials with $B_N$ type weight function, the detailed derivation of $SO(2N)$ case will be attached to Appendix \ref{app3}, while the $Sp(2N)$, $SO(2N+1)$ cases follow similar procedures. In this paper, we will investigate both $I_{SO(2N+1)}(u,q,t)$ and $I'_{B_N}(u,q,t)$, adopting the notational convention $I'_{SO(2N+1)}(u,q,t):=I'_{B_N}(u,q,t)$ for conceptual clarity.
The Koornwinder weight function naturally induces a scalar product for Koornwinder polynomials $K_{ \lambda}(x; q, t; t_0,t_1,t_2,t_3)$.  The inherent inhomogeneity of these polynomials presents a significant obstacle to obtaining results as elegant as those in the $U(N)$ case. Nevertheless, for both the $BCD_N$ type deformed Schur indices and our newly defined $I'_{SO(2N+1)}$, by transforming Koornwinder polynomials into Macdonald polynomials yield a  combinatorial like results. However, this procedure requires term-by-term computations that are inherently intricate. To do this, we need a theorem of Koornwinder polynomials\cite{MR4259867}:
\begin{theorem} \label{thm2}
\begin{align*}
\sum_\text{$\lambda$ is even}b^{\mathrm{oa}}_{\lambda;m}(q,t)P_\lambda(x;q,t)&=(x_1 \cdots x_n)^m K_{m^n}(x; q, t; +\sqrt{q},-\sqrt{q},+\sqrt{qt},-\sqrt{qt}), \\[8pt]
\sum_{\lambda\subset m^n}b^{\mathrm{el}}_{\lambda;2m}(q,t)P_\lambda(x;q,t)&=(x_1 \cdots x_n)^m K_{m^n }(x; q, t; -1,+t,-\sqrt{q},+\sqrt{q}),\\[8pt]
\sum_{\substack{\text{$\lambda\subset m^n$} \\[3pt] \text{$m_i(\lambda)$ is even}}}b^{\mathrm{ol}}_{\lambda;m}(q,t)P_\lambda(x;q,t)&=(x_1 \cdots x_n)^m K_{m^n }(x; q, t; -1,+1,-\sqrt{q},+\sqrt{q}),\\[8pt]
\sum_{\lambda\subset m^n}a^{odd(\lambda')}b^{\mathrm{el}}_{\lambda;2m}(q,t)P_\lambda(x;q,t)&=\prod_{i=1}^n\frac{(tax_i,q)_\infty}{(ax_i,q)_\infty}(x_1 \cdots x_n)^m K_{m^n }(x; q, t; -1,+1,-\sqrt{q},+\sqrt{q}),
\end{align*}
\end{theorem}
\noindent where $m_i(\lambda)=\#\{\lambda_j|\lambda_j=i\}$, $odd(\lambda)=\#\{\lambda_i|\text{$\lambda_i$ is odd}\}$. Similar summations over subsets of Young tableaux also appeared in \cite{Sei:2023fjk,du2023schurindicesmathcaln4superyangmills}. The different type coefficients $b_{\lambda}$ are defined as:
\begin{align*}
        &b_\lambda(s;q,t) := \frac{1 - q^{a_\lambda(s)} t^{l_\lambda(s)+1}}{1 - q^{a_\lambda(s)+1} t^{l_\lambda(s)}},\quad\quad\quad\quad b_{\lambda}(q,t) := \prod_{s\in\lambda} b_{\lambda}(s;q,t),\\
        &b_\lambda^{\mathrm{oa}}(q,t) := \prod_{\substack{s\in\lambda\\a(s)\text{ odd}}} b_\lambda(s;q,t),\ b_\lambda^{\mathrm{el}}(q,t) := \prod_{\substack{s\in\lambda\\l(s)\text{ even}}} b_\lambda(s;q,t),\ b_\lambda^{\mathrm{ol}}(q,t) :=\prod_{\substack{s \in \lambda \\ l(s)\ \text{odd}}}\frac{1 - q^{a(s)} t^{l(s)}}{1 - q^{a(s)+1} t^{l(s)-1}},\\[7pt]
        &b^{\mathrm{oa}}_{\lambda;m}(q,t) := b^{\mathrm{oa}}_{\lambda}(q,t) \prod_{\substack{s\in\lambda\\ a'(s)\text{ odd}}}\frac{1 - q^{2m-a'(s)+1} t^{l'(s)}}{1 - q^{2m-a'(s)} t^{l'(s)+1}},\\
        &b_{\lambda;m}^{\mathrm{ol}}(q,t) := b_\lambda^{\mathrm{ol}}(q,t)\prod_{\substack{s \in \lambda \\ l'(s) \text{ odd}}} \frac{1 - q^{m-a'(s)} t^{l'(s)-1}}{1 - q^{m-a'(s)-1} t^{l'(s)}}, \\
        &b_{\lambda;m}^{\mathrm{el}}(q,t) := b_{\lambda}^{\mathrm{el}}(q,t) \prod_{\substack{s \in \lambda \\ l'(s) \text{ even}}} \frac{1 - q^{m-a'(s)} t^{l'(s)}}{1 - q^{m-a'(s)-1} t^{l'(s)+1}},
\end{align*}
where $a(s)$ is called arm-length, $l(s)$ is called leg-length, their exact definitions are provided in Appendix \ref{appendixA}. 
In our paper, we only focus on $m \to \infty$ limit, due to $|q|<1$, we have:
\begin{equation*}
    b^{\mathrm{oa}}_{\lambda;\infty}(q,t)=b^{\mathrm{oa}}_{\lambda}(q,t),\quad\quad b^{\mathrm{ol}}_{\lambda;\infty}(q,t)=b^{\mathrm{ol}}_{\lambda}(q,t),\quad\quad b^{\mathrm{el}}_{\lambda;\infty}(q,t)=b^{\mathrm{el}}_{\lambda}(q,t), 
\end{equation*}
for $|q|>1$ these can be defined by analytic continuation. And we see that the parameter specialization  first line in Theorem \ref{thm2} is different with it in the $Sp(2N)$ deformed Schur index, but in the limit $q=t$ they will coincide with each others, so we contain it in this paper and will be used in our discussion of further limit in Sec \ref{sec5}.
But we can conjecture similar identity for Koornwinder polynomials in $Sp(2N)$ specialization:
\begin{align}
    \sum_\text{$|\lambda|$ is even}b^{\mathrm{e}}_{\lambda;m\to \infty}(q,t)P_\lambda(x;q,t)&=(x_1 \cdots x_n)^{m\to \infty} K_{(m\to \infty)^n}(x; q, t; +\sqrt{t},-\sqrt{t},+\sqrt{qt},-\sqrt{qt}).
\end{align}
After some calculation, we can get exact value of $b^{\mathrm{e}}_{\lambda}(q,t)$ corresponding to low-rank partitions:
\begin{align*}
    &b^{\mathrm{e}}_{[0]}(q,t)=1,\quad\quad b^{\mathrm{e}}_{[1,1]}(q,t)=\frac{(1-t)(q-t)}{(qt-1)(1-q)},\quad\quad b^{\mathrm{e}}_{[2]}(q,t)=\frac{(1-t)}{(1-q)},\\
    &b^{\mathrm{e}}_{[1,1,1,1]}(q,t)=\frac{(1-t)(q-t)t(1-t^3)}{(qt-1)(1-q)(1-qt^3)},\quad\quad b^{\mathrm{e}}_{[2,1,1]}(q,t)=\frac{(1-t)^2(t-q)(1+qt^2)}{(qt-1)^2(1-q)(1-q^2t^2)},\\
    &b^{\mathrm{e}}_{[3,1]}(q,t)=\frac{(1-t)^2(t-q)q}{(1-q)^2(1-q^3t)},\quad\quad b^{\mathrm{e}}_{[4]}(q,t)=\frac{(1-t)(1-qt)}{(1-q)(1-q^2)}.
\end{align*}
Based on the all contexts above, we can reformulate the integral expressions for the  $BCD_N$ type deformed Schur indices as follows: 
\begin{align}\label{KoornwidnerSp(2N)aftertheorem2}
I_{Sp(2N)} (u,q,t)  &=
\begin{aligned}[t]
    &\mathcal{X}_N \sum_{\lambda}\sum_{\text{$|\mu|$,$|\nu|$ is even}}u^{|\lambda|+\frac{|\mu|+|\nu|}{2}}b_{\lambda}(q,t)b^{\mathrm{e}}_{\mu}(q,t)b^{\mathrm{e}}_{\nu}(q,t)\\
&\times \oint_{\mathbb{T}^N} \ \Delta_N(\mathbf{x};q,t;\pm\sqrt{t},\pm\sqrt{qt}) \Delta_N(\mathbf{x^{-1}};q,t;\pm\sqrt{t},\pm\sqrt{qt}) \\
&\times P_{\lambda}(\mathbf{x}; q, t) P_{\mu}(\mathbf{x}; q, t)P_{\lambda}(\mathbf{x}^{-1}; q, t)P_{\nu}(\mathbf{x}^{-1}; q, t)\prod_{j=1}^N\frac{dx_j}{2\pi ix_j},
\end{aligned}\\[2ex]
\label{KoornwidnerSO(2N+1)âaftertheorem2}
I'_{SO(2N+1)} (u,q,t)  &= 
\begin{aligned}[t]
&\mathcal{X}_N \sum_{\lambda}\sum_{\mu,\nu}u^{|\lambda|+\frac{|\mu|+|\nu|}{2}}b_{\lambda}(q,t)b^{\mathrm{el}}_{\mu}(q,t)b^{\mathrm{el}}_{\nu}(q,t)\\
&\times \oint_{\mathbb{T}^N} \ \Delta_N(\mathbf{x};q,t;-1,+t,-\sqrt{q},+\sqrt{q}) \Delta_N(\mathbf{x^{-1}};q,t;-1,+t,-\sqrt{q},+\sqrt{q}) \\
&\times P_{\lambda}(\mathbf{x}; q, t) P_{\mu}(\mathbf{x}; q, t)P_{\lambda}(\mathbf{x}^{-1}; q, t)P_{\nu}(\mathbf{x}^{-1}; q, t)\prod_{j=1}^N\frac{dx_j}{2\pi ix_j},
\end{aligned}\\[2ex]
\label{KoornwidnerSO(2N)aftertheorem2}
I_{SO(2N)} (u,q,t)  &=
\begin{aligned}[t]
&\mathcal{X}_N \sum_{\lambda}\sum_{\text{$m_i(\mu)$,$m_i(\nu)$is even}}u^{|\lambda|+\frac{|\mu|+|\nu|}{2}}b_{\lambda}(q,t)b^{\mathrm{ol}}_{\mu}(q,t)b^{\mathrm{ol}}_{\nu}(q,t)\\
&\times \oint_{\mathbb{T}^N} \ \Delta_N(\mathbf{x};q,t;-1,+1,-\sqrt{q},+\sqrt{q}) \Delta_N(\mathbf{x^{-1}};q,t; -1,+1,-\sqrt{q},+\sqrt{q}) \\
&\times P_{\lambda}(\mathbf{x}; q, t) P_{\mu}(\mathbf{x}; q, t)P_{\lambda}(\mathbf{x}^{-1}; q, t)P_{\nu}(\mathbf{x}^{-1}; q, t)\prod_{j=1}^N\frac{dx_j}{2\pi ix_j},
\end{aligned}\\[2ex]
\label{KoornwidnerSO(2N+1)aftertheorem2}
I_{SO(2N+1)} (u,q,t)  &= 
\begin{aligned}[t]
&\mathcal{X}_N \sum_{\lambda}\sum_{\mu,\nu}u^{|\lambda|+\frac{|\mu|+|\nu|+odd(\mu')+odd(\nu')}{2}}b_{\lambda}(q,t)b^{\mathrm{el}}_{\mu}(q,t)b^{\mathrm{el}}_{\nu}(q,t)\\
&\times \oint_{\mathbb{T}^N} \ \Delta_N(\mathbf{x};q,t;-1,+t,-\sqrt{q},+\sqrt{q}) \Delta_N(\mathbf{x^{-1}};q,t;-1,+t,-\sqrt{q},+\sqrt{q}) \\
&\times P_{\lambda}(\mathbf{x}; q, t) P_{\mu}(\mathbf{x}; q, t)P_{\lambda}(\mathbf{x}^{-1}; q, t)P_{\nu}(\mathbf{x}^{-1}; q, t)\prod_{j=1}^N\frac{dx_j}{2\pi ix_j}.
\end{aligned}
\end{align}
An alternative approach for deriving the exact deformed Schur index for $SO(2N+1)$ gauge groups involves expanding the additional $q$-Pochhammer symbol $\prod_{i=1}^n \frac{(tux_i;q)_\infty}{(ux_i;q)_\infty}$ in terms of Macdonald polynomials corresponding to special partitions\cite{macdonald1998symmetric}\cite{noumi2023macdonald}:
\begin{align}\label{expansion1}
G(x;u) = \prod_{i=1}^n \frac{(tux_i;q)_\infty}{(x_iu;q)_\infty} = \sum_{l=0}^\infty \frac{(t;q)_l}{(q;q)_l} P_{[l]}(\mathbf{x};q,t) u^l,
\end{align}
denote $w_l(q,t):=\frac{(t;q)_l}{(q;q)_l}$, then the deformed index $I_{SO(2N+1)}(u,q,t)$ becomes: 
\begin{align}
\label{KoornwidnerSO(2N+1)aftertheorem2andspecexpans}
    \begin{split}
    I_{SO(2N+1)} &(u,q,t)  = \mathcal{X}_N \sum_{l=0}^\infty\sum_{\lambda}\sum_{\text{$m_i(\mu)$,$m_i(\nu)$is even}}u^{|\lambda|+l+\frac{|\mu|+|\nu|}{2}}b_{\lambda}(q,t)b^{\mathrm{ol}}_{\mu}(q,t)b^{\mathrm{ol}}_{\nu}(q,t)w_l(q,t)\\
&\times \oint_{\mathbb{T}^N} \ \Delta_N(\mathbf{x};q,t;-1,+t,-\sqrt{q},+\sqrt{q}) \Delta_N(\mathbf{x^{-1}};q,t; -1,+t,-\sqrt{q},+\sqrt{q}) \\
&\times P_{\lambda}(\mathbf{x}; q, t) P_{\mu}(\mathbf{x}; q, t)P_{[l]}(\mathbf{x}; q, t)P_{\lambda}(\mathbf{x}^{-1}; q, t)P_{\nu}(\mathbf{x}^{-1}; q, t)P_{[l]}(\mathbf{x}^{-1}; q, t)\prod_{j=1}^N\frac{dx_j}{2\pi ix_j}.
    \end{split}
\end{align}
The combinatorial evaluations of these integrals requires careful treatment of the Koornwinder weight function. Through its decomposition into two distinct components, one component identical to the weight function (\ref{weightfunction}), remaining component is expressible through monomial symmetric functions. We establish that all terms in the resulting summation manifest as linear combinations of normalization factors (\ref{normfactor}), each corresponding to distinct partitions $\lambda$. The complete framework and detailed calculation will be systematically developed in  Section \ref{sec3}.

\section{Expansion in Macdonald normalization constants}
\label{sec3}
This section presents the computational framework for evaluating the deformed Schur indices through expansion of $q$-Pochhammer symbols in the integral representation. Our approach specifically excludes terms already accounted for in the $U(N)$ case (\ref{weightfunction}), (\ref{macintegrand}), then restructure it into monomial symmetric functions,

Following the previous section \ref{sec2}, we first expand the index in the $u$ fugacity parameter.  In the $U(N)$ case in \cite{hatsuda2025deformedschurindicesmacdonald}, the expansion coefficients are very simply related to the normalization constants of the Macdonald polynomial, and one obtains  nice closed formulas. Our situation is more complicated, we need to use some combinatorial rules of symmetric functions to obtain a further expansion in terms of these normalization constants. Unlike \cite{hatsuda2025deformedschurindicesmacdonald}, the results in this section cannot be expanded in small fugacity parameters  because higher order normalization constant terms may contribute to the low order terms in the expansion of fugacity parameters. Nevertheless, we hope the results would be still useful for further studies.

We start with integrals (\ref{Cdsi}), (\ref{Bdsi}), (\ref{Ddsi}) without the help of Theorem.\ref{thm1}, Theorem.\ref{thm2} and Koornwinder polynomials, employing some properties of $q$-Pochhammer symbols:
\begin{align*}
\begin{split}
   &(a,q)_0 = 1,\quad (a; q)_n = \frac{(a; q)_\infty}{(a q^n; q)_\infty},\quad \frac{(ax; q)_\infty}{(x; q)_\infty} = \sum_{n=0}^\infty \frac{(a; q)_n}{(q; q)_n} x^n,\\[2pt]
   &(a;q^{-1})_n=(a^{-1};q)_n (-a)^n q^{\binom{n}{2}},\quad (aq^k;q)_{n-k}=\frac{(a; q)_n}{(q; q)_k},\\
   &\frac{(a; q)_\infty(tua; q)_\infty}{(ta; q)_\infty(ua; q)_\infty}=\sum_{k=0}^{\infty}\Bigg[\sum_{n=0}^k\frac{(t; q^{-1})_n(t; q)_{k-n}}{(q; q)_n(q; q)_{k-n}}(-1)^nu^{k-n}q^{-\binom{n}{2}}\Bigg]a^k,\\[1ex]
\end{split}   
\end{align*}
define 
\begin{align*}
\begin{split}
&c_k(u;q,t):=\sum_{n=0}^k\frac{(t; q^{-1})_n(t; q)_{k-n}}{(q; q)_n(q; q)_{k-n}}(-1)^nu^{k-n}q^{-\binom{n}{2}},\\
&c_{n_1,n_2,\dots,n_m}(u;q,t):=c_{n_1}(u;q,t)c_{n_2}(u;q,t)\cdots c_{n_m}(u;q,t),\\
&c_{n_1,n_2,\dots,n_i,0,\dots,0}(u;q,t)=c_{n_1,n_2,\dots,n_i}(u;q,t).\\
\end{split}   
\end{align*}

For SO(2N), the integrand of the deformed Schur index becomes:
\begin{align}
    \begin{split}
      w(\mathbf{x}) &\sum_{\ell(\lambda) \leq N} u^{|\lambda|} b_\lambda P_\lambda(\mathbf{x};q,t) P_\lambda(\mathbf{x}^{-1};q,t)\\
      &\prod_{1 \leq i < j \leq N} \left[\sum_{k_{ij}^+=0}^{\infty}c_{k_{ij}^+}(u;q,t)(x_i x_j)^{k_{ij}^+}\right] \prod_{1 \leq i < j \leq N} \left[\sum_{k_{ij}^-=0}^{\infty}c_{k_{ij}^-}(u;q,t)(x_i x_j)^{-k_{ij}^-}\right].
    \end{split}
\end{align} 
The simplest case is $SO(4)$ ($N=2$) , this can be written as:
\begin{align}
    \begin{split}
      w(\mathbf{x}) &\sum_{\ell(\lambda) \leq 2} u^{|\lambda|} b_\lambda P_\lambda(\mathbf{x};q,t) P_\lambda(\mathbf{x}^{-1};q,t)\\
      &\times \left[\sum_{n^+=0}^{\infty}\left(c_{n^+}(u,q,t)\right)(x_1 x_2)^{n^+}\right]\times \left[\sum_{n^-=0}^{\infty}\left(c_{n^-}(u,q,t)\right)(x_1 x_2)^{-n^-}\right],
    \end{split}
\end{align}
using the Pieri rules:
\begin{align}
P_\mu(x_1, \dots, x_n; q, t) \cdot x_1 \cdots x_n \nonumber = P_{\mu + 1^n}(x_1, \dots, x_n; q, t),
\end{align}
together with the orthogonality defined by the integral (\ref{scpro}), the deformed Schur index for $SO(4)$ can be expanded as:
\begin{align*}
I_{SO(4)} (u,q,t)  = \mathcal{X}_2  \sum_{\ell(\lambda) \leq 2}\sum_{n} u^{|\lambda|} b_\lambda c^2_{n}(u,q,t) \mathcal{N}_{\lambda+n^2,2}.
\end{align*}
And $\mathcal{N}_{\lambda+n^N,N}=\mathcal{N}_{\lambda,N}$ for any $n \in \mathrm{Z}_+$, 
 follows directly from the definition of normalization factor, so
\begin{equation}
    I_{SO(4)} (u,q,t) =\frac{1}{2}\sum_{n}c^2_{n}(u,q,t)I_{SU(2)} (u,q,t).
\end{equation}
For $SO(6)$ ($N=3$) case:
\begin{align*}
    \begin{split}
      w(\mathbf{x})& \sum_{\ell(\lambda) \leq N} u^{|\lambda|} b_\lambda P_\lambda(\mathbf{x};q,t) P_\lambda(\mathbf{x}^{-1};q,t)\\
      &\times \left[\sum_{{n_{12}^+,n_{13}^+,n_{23}^+=0}}^{\infty}c_{n_{12}^+,n_{13}^+,n_{23}^+}\mathbf{x}^{A(n_{12}^+,n_{13}^+,n_{23}^+)}\right] \left[\sum_{{n_{12}^-,n_{13}^-,n_{23}^-=0}}^{\infty}c_{{n_{12}^-,n_{13}^-,n_{23}^-}}(\mathbf{x}^{-1})^{A(n_{12}^-,n_{13}^-,n_{23}^-)}\right],\\
    \end{split}
\end{align*}
where
\begin{align*}
    \begin{split}
       &\mathbf{x}^{A(n_{12}^+,n_{13}^+,n_{23}^+)}:=(x_1)^{n_{12}^++n_{13}^+}(x_2)^{n_{12}^++n_{23}^+}(x_3)^{n_{13}^++n_{23}^+},\\
       &(\mathbf{x}^{-1})^{A(n_{12}^-,n_{13}^-,n_{23}^-)}:=(x^{-1}_1)^{n_{12}^-+n_{13}^-}(x^{-1}_2)^{n_{12}^-+n_{23}^-}(x^{-1}_3)^{n_{13}^-+n_{23}^-}.
    \end{split}
\end{align*}
Define $a_1:=n_{12}^++n_{13}^+$, $a_2:=n_{12}^++n_{23}^+$, $a_3:=n_{13}^++n_{23}^+$, the coefficient function $c_{n_{12}^+,n_{13}^+,n_{23}^+}$ is invariant under the natural action of $S_3$ permuting ${n_{12}^+,n_{13}^+,n_{23}^+}$. For monomial symmetric function construction, we require the induced action $S_3'$ on ${a_1,a_2,a_3}$. And this action can be induced by the original $S_3$. The groups $S_3$ and $S_3'$ are isomorphic to each other$\footnote{The transposition $(a_1,a_2)\in S'_3$ is induced by the transposition $(n_{13}^+,n_{23}^+)\in S_3$, so as other fundamental transpositions in $S'_3$, so $S'_3\cong S_3$. }$. In our analysis, we begin by considering the full symmetric group $S_3$ and summing over all terms sharing identical coefficients $c_{n_{12}^+,n_{13}^+,n_{23}^+}(u,q,t)$, computational redundancy arises when any of the indices ${n_{12}^+,n_{13}^+,n_{23}^+}$  coincide, situation is same when summing over $S'_3$ acting on $a_1,a_2,a_3$. Redundancy in summing over the $S_3$ orbits will contribute an overall factor, in summing over $S_3'$ generates partition-specific factors dependent on $\lambda$, for general $N$. For arbitrary $N$ these two redundancy factors may differ, but in $N=3$ case, they become identical and consequently cancel in our calculations\footnote{$a_1=a_2$ if and only if $ n_{13}^+=n_{23}^+$, which is same for other pair of numbers in two space.}. After doing this the deform Schur index for $SO(6)$ can be written as:
\begin{align*}
I_{SO(6)} (u,q,t)  = \mathcal{X}_3  \sum_{\ell(\lambda) \leq 3} u^{|\lambda|} b_\lambda &\sum_{{n_{12}^+\geq n_{13}^+\geq n_{23}^+}}^{\infty}\sum_{{n_{12}^-\geq n_{13}^-\geq n_{23}^-}}^{\infty}\frac{c_{n_{12}^+,n_{13}^+,n_{23}^+}}{r_{[n_{12}^+,n_{13}^+,n_{23}^+]}}\frac{c_{n_{12}^-,n_{13}^-,n_{23}^-}}{r_{[n_{12}^-,n_{13}^-,n_{23}^-]}} \\
&\left\langle r_{\mu}m_{\mu}(\mathbf{x})P_{\lambda}(\mathbf{x}), r_{\nu}m_{\nu}(\mathbf{x})P_{\lambda}(\mathbf{x})\right \rangle,
\end{align*}
where:
\begin{align*}
   &\mu=[n_{12}^++n_{13}^+,n_{12}^++n_{23}^+,n_{13}^++n_{23}^+],\quad \nu=[n_{12}^-+n_{13}^-,n_{12}^-+n_{23}^-,n_{13}^-+n_{23}^-],\\
   &r_{\lambda}:=\prod_{i=1}^N \sqrt[m_{\lambda_i}(\lambda)]{m_{\lambda_i}(\lambda)!},
\end{align*}
and easily checked that $r_{\mu}=r_{[n_{12}^+,n_{13}^+,n_{23}^+]}$ and $r_{\nu}=r_{[n_{12}^-,n_{13}^-,n_{23}^-]}$ for any ${n_{12}^+,n_{13}^+,n_{23}^+}$ and ${n_{12}^-,n_{13}^-,n_{23}^-}$, so we can simplify above results to :
\begin{align}
\begin{split}
I_{SO(6)} (u,q,t)  = \mathcal{X}_3  \sum_{\ell(\lambda) \leq 3} u^{|\lambda|} b_\lambda &\sum_{{n_{12}^+\geq n_{13}^+\geq n_{23}^+}}^{\infty}\sum_{{n_{12}^-\geq n_{13}^-\geq n_{23}^-}}^{\infty}c_{n_{12}^+,n_{13}^+,n_{23}^+} c_{n_{12}^-,n_{13}^-,n_{23}^-}\\
&\times\left\langle m_{\mu}(\mathbf{x})P_{\lambda}(\mathbf{x}), m_{\nu}(\mathbf{x})P_{\lambda}(\mathbf{x})\right \rangle.
\end{split}
\end{align}
Because $m_{\lambda}(\mathbf{x})$, $P_{\lambda}(\mathbf{x})$ are both homogeneous symmetric polynomials with degree $|\lambda|$, so only $n_{12}^++n_{13}^++n_{23}^+=n_{12}^-+n_{13}^-+n_{23}^-$ will contribute, The symmetric function product appearing in the inner product can be systematically decomposed through expansion in the basis of Macdonald polynomials, and some results which are essential in this paper will be listed in Appendix \ref{appendixB}. We proceed to expand the deformed Schur index using the normalization factor defined in (\ref{normfactor}). For concreteness, we present the leading-order terms corresponding to low-rank partitions:
\begin{align}
\begin{split}
\label{SO(6)}
&I_{SO(6)} (u,q,t)\\
&= \mathcal{X}_3 \Bigg[\mathcal{N}_{[0],3}+u  b_{[1]}\mathcal{N}_{[1],3}+u^2b_{[2]}\mathcal{N}_{[2],3}+\Bigg(u^2b_{[1,1]}+c_{1,0,0}^2\Bigg)\mathcal{N}_{[1,1],3}+\Bigg(u^3 b_{[2,1]}+uc_{1,0,0}^2\frac{1-t}{1-q}\Bigg)\mathcal{N}_{[2,1],3}\\
&\quad\quad+u^3 b_{[3]}\mathcal{N}_{[3],3}+\Bigg(u^3 b_{[1,1,1]}+uc_{1,0,0}^2\frac{1-t}{1-q}\left( \frac{-q t^2 - q t + t^2 - q + t + 1}{-q t^2 + 1} \right)^2\Bigg)\mathcal{N}_{[1,1,1],3}\\
&\quad\quad+\Bigg( u^4b_{[2,1,1]}+u^2c_{1,0,0}^2\Bigg(b_{[1,1]}(\left( \frac{q t + q - t - 1}{q t - 1}\right)^2+b_{[2]}\left( \frac{q^3 t^3 - q t^3 - q^2 + 1}{q^3 t^3 - q^2 t^2 - qt + 1} \right)^2\Bigg)+c_{1,1,0}^2\\
&\quad\quad\quad\quad+c_{2,0,0}^2\left(\frac{-qt+q-t+1}{qt-1}\right)^2+2c_{1,1,0}c_{2,0,0}\frac{-qt+q-t+1}{qt-1}\Bigg)\mathcal{N}_{[2,1,1],3}+u^4b_{[4]}\mathcal{N}_{[4],3}\\
&\quad\quad+ \Bigg(u^4b_{[3,1]}+u^2c_{1,0,0}^2b_{[2]}\Bigg)\mathcal{N}_{[3,1],3}+ \Bigg(u^4b_{[2,2]}+u^2c_{1,0,0}^2b_{[1,1]}+c_{2,0,0}^2\Bigg)\mathcal{N}_{[2,2],3}+\mathcal{O}\Bigg(\mathcal{N}_{|\lambda|>4,3}\Bigg)\Bigg],
\end{split}
\end{align}
$\scalebox{0.8}{$\mathcal{O}\Bigg(\mathcal{N}_{|\lambda|>4,3}\Bigg)$}$ here means non-trivial summation $\sum\limits_{|\lambda|>4, \ell(\lambda)\leq 3}f_{\lambda}(u,q,t)\mathcal{N}_{|\lambda|>4,3}$ and the coefficient $f_{\lambda}(u,q,t)$ will be much more complicated for large $\lambda$, and
\begin{align*}
&b_{[1]}=\frac{1-t}{1-q},\ b_{[2]}=\frac{(1-qt)(1-t)}{(1-q^2)(1-q)},\ b_{[1,1]}=\frac{(1-t)(1-t^2)}{(1-q)(1-qt)},\ b_{[1,1,1]}=\frac{(1-t)(1-t^2)(1-t^3)}{(1-q)(1-qt)(1-qt^2)},\\[1ex]
&b_{[2,1]}=\frac{(1-qt)(1-t)(1-t^2)}{(1-q^2)(1-q)(1-qt)},\ b_{[3]}=\frac{(1-q^2t)(1-qt)(1-t)}{(1-q^3)(1-q^2)(1-q)},\ b_{[3,1]}=\frac{(1-q^2t)(1-t)(1-t^2)}{(1-q^3)(1-q^2)(1-q)},\\[1ex]
&b_{[1,1,1,1]}=\frac{(1-t)(1-t^2)(1-t^3)(1-t^4)}{(1-q)(1-qt)(1-qt^2)(1-qt^3)},\quad b_{[2,1,1]}=\frac{(1-t)(1-t^2)(1-t^3)}{(1-q^2)(1-q)(1-qt^2)},\\[1ex]
&b_{[2,2]}=\frac{(1-t)(1-t^2)(1-qt^2)}{(1-q^2)(1-q)(1-q^2t)},\quad b_{[4]}=\frac{(1-q^3t)(1-q^2t)(1-qt)(1-t)}{(1-q^4)(1-q^3)(1-q^2)(1-q)},\\[1ex]
&c_{1,0,0}=\frac{(1-t)}{1-q}(u-1),\quad c_{1,1,0}=\frac{(1-t)^2}{(1-q)^2}(u-1)^2,\\[1ex]
&c_{2,0,0}=\frac{(1-t)(1-qt)}{(1-q)(1-q^2)}u^2-\frac{(1-t)^2}{(1-q)^2}u+\frac{(1-t)(1-tq^{-1})}{(1-q)(1-q^2)}q.
\end{align*}

For the general case when $N \geq 4$, our expansion involves coefficients $c_{n_{12}^+, ...,n_{N-1,N}^+}$ that remain invariant under the action of the symmetric group $S_{\frac{N(N-1)}{2}}$. We define the auxiliary variables: $a_1:=n_{12}^++n_{13}^++\cdots +n_{1,N}^+, a_2:=n_{12}^++n_{23}^+\cdots +n_{2,N}^+,\cdots, a_N:=n_{1,N}^++n_{2N}^++\cdots +n_{N-1,N}^+$. Two significant distinctions compared to the  $N=2,3$ case, the first is that for a given configuration ${{n_{12}^+,n_{13}^+,\cdots ,n_{N-1,N}^+}}$ may correspond to multiple distinct partitions, consequently, multiple types of monomial symmetric functions appear in the expansion. The symmetric group $S'_N$ act on $a_1,a_2,\cdots,a_N$ which can be constructed by its induced action now is just a subgroup of $S_{\frac{N(N-1)}{2}}$, and the redundancies of summing over the orbits under the action of these two groups are not necessarily the same anymore, but fortunately, our construction of monomial symmetric function still works but we should take care about an additional factor for different monomial symmetric functions, and this factor admits an interpretation in terms of  the structure of symmetric group and its orbits.  Omitting the detailed derivation we present the final expression directly for $CD_N$ type deformed Schur indices: 
\begin{align}
\begin{split}
I_{SO(2N)} (u,q,t)  &= \mathcal{X}_N  \sum_{\ell(\lambda) \leq N} u^{|\lambda|} b_\lambda \\
&\times \sum_{{n_{12}^+\geq n_{13}^+\geq\cdots \geq n_{N-1,N}^+}}^{\infty}\sum_{{n_{12}^-\geq n_{13}^-\geq\cdots \geq n_{N-1,N}^-}}^{\infty}c_{n_{12}^+,n_{13}^+,\cdots ,n_{N-1,N}^+} c_{{n_{12}^-,n_{13}^-,\cdots ,n_{N-1,N}^-}}\\
&\times\sideset{}{^{\{n^+\}}}\sum_{\mu}\sideset{}{^{\{n^-\}}}\sum_{\nu}\left\langle n_{\mu}m_{\mu}(\mathbf{x})P_{\lambda}(\mathbf{x}), n_{\nu}m_{\nu}(\mathbf{x})P_{\lambda}(\mathbf{x})\right \rangle.
\end{split}
\end{align}
The summation $\scalebox{0.7}{$\sideset{}{^{\{n^+\}}}\sum\limits_{\mu}$}$ $ $ means summing over all distinct partitions for a given configuration $\{({n_{12}^{'+},n_{13}^{'+},\cdots ,n_{N-1,N}^{'+}})|({n_{12}^{'+},n_{13}^{'+},\cdots ,n_{N-1,N}^{'+}})=\sigma ({n_{12}^{+},n_{13}^{+},\cdots ,n_{N-1,N}^{+}}),\ \forall\sigma \in S_{{N(N-1)}/{2}}\}$. And the factor $n_{\mu}$ can be interpreted as follow: when some of numbers ${n_{12}^+,n_{13}^+,\cdots ,n_{N-1,N}^+}$ are same, the redundancy will be described by a subgroup $G$, we can do coset decomposition respect to this group. For  $\mu$ associated with the configuration ${n_{12}^+,n_{13}^+,\cdots ,n_{N-1,N}^+}$, the redundancy group $G_{\mu}'\in S_{\frac{N(N-1)}{2}}$ can be directly determined. Every elements in $G_{\mu}'$ belongs to coset of $G$ in $S_{\frac{N(N-1)}{2}}$, then the factor $n_{\mu}$ is just the numbers of the cosets that $G_{\mu}'$ intersect, i.e. the factor $n_{\mu}$ encodes the overlap between the $G'_{\mu}$ and $G$-coset structure. 

For example, $SO(8)$ deformed Schur index: 
\begin{align*}
&I_{SO(8)} (u,q,t)=\\
& \mathcal{X}_3 \Bigg[\mathcal{N}_{[0],4}+u  b_{[1]}\mathcal{N}_{[1],4}+u^2b_{[2]}\mathcal{N}_{[2],4}+\Bigg(u^2b_{[1,1]}+c_{1}^2\Bigg)\mathcal{N}_{[1,1],4}+\Bigg(u^3 b_{[2,1]}+uc_{1}^2\frac{1-t}{1-q}\Bigg)\mathcal{N}_{[2,1],4}\\
&\quad+u^3 b_{[3]}\mathcal{N}_{[3],4}+\Bigg(u^3 b_{[1,1,1]}+uc_{1}^2\frac{1-t}{1-q}\left( \frac{-q t^2 - q t + t^2 - q + t + 1}{-q t^2 + 1} \right)^2\Bigg)\mathcal{N}_{[1,1,1],4}\\
&\quad+\Bigg( u^4b_{[2,1,1]}+u^2c_{1}^2\Bigg(b_{[1,1]}(\left( \frac{q t + q - t - 1}{q t - 1}\right)^2+b_{[2]}\left( \frac{q^3 t^3 - q t^3 - q^2 + 1}{q^3 t^3 - q^2 t^2 - qt + 1} \right)^2\Bigg)+c_{1,1}^2\\
&\quad\quad\quad+c_{2}^2\left(\frac{-qt+q-t+1}{qt-1}\right)^2+2c_{1,1}c_{2}\frac{-qt+q-t+1}{qt-1}\Bigg)\mathcal{N}_{[2,1,1],4}+u^4b_{[4]}\mathcal{N}_{[4],4}\\
&\quad+\Bigg(u^4b_{[3,1]}+u^2c_{1}^2b_{[2]}\Bigg)\mathcal{N}_{[3,1],4}+ \Bigg(u^4b_{[2,2]}+u^2c_{1}^2b_{[1,1]}+c_{2}^2\Bigg)\mathcal{N}_{[2,2],4}\\
&\quad+\Bigg(u^2c_{1}^2b_{[1,1]}B^2+u^4b_{[1,1,1,1]}+\Bigg(c_{1,1}\Bigg(A+3\Bigg)+c_{2}B\Bigg)^2\Bigg)\mathcal{N}_{[1,1,1,1],4}+\mathcal{O}\Bigg(\mathcal{N}_{|\lambda|>4,4}\Bigg)\Bigg],
\end{align*}
where
\begin{flalign*}
&c_{1}=\frac{(1-t)}{1-q}(u-1),\  c_{1,1}=\frac{(1-t)^2}{(1-q)^2}(u-1)^2,\\
&c_{2}=\frac{(1-t)(1-qt)}{(1-q)(1-q^2)}u^2-\frac{(1-t)^2}{(1-q)^2}u+\frac{(1-t)(q-t)}{(1-q)(1-q^2)},&\\[1ex]
&A=\frac{-3 q t^3 + q t^2 - t^3 + q t - t^2 + q - t + 3}{q t^3 - 1},&\\[1ex]
&B=\frac{q^2 t^5 - q^2 t^4 + q t^5 - q^2 t^2 - q t^3 + t^4 + q^2 t - q t^2 - t^3 + q - t + 1}{q^2 t^5 - q t^3 - q t^2 + 1}.&
\end{flalign*}

The deformed Schur index for $Sp(2N)$ can be obtained directly from the $SO(2N)$ case by taking the following substitutions:
\begin{align*}
    \begin{split}
       c_{n_{12}^+,n_{13}^+,\cdots ,n_{N-1,N}^+} &\to   c_{n_{12}^+,n_{13}^+,\cdots ,n_{N-1,N}^+,n_1^+,n_2^+,\cdots, n_N^+},\\
       A(n_{12}^+,n_{13}^+,\cdots ,n_{N-1,N}^+)&\to A'(n_{12}^+,n_{13}^+,\cdots ,n_{N-1,N}^+,n_1^+,n_2^+,\cdots, n_N^+),\\
       A(n_{12}^+,n_{13}^+,\cdots ,n_{N-1,N}^+)&=(n_{12}^++n_{13}^++\cdots+n_{1,N}^+,\cdots,n_{1,N}^++n_{2,N}^++\cdots+n_{N-1,N}^+),\\
       A'(n_{12}^+,\cdots ,n_{N-1,N}^+,n_1^+,\cdots, n_N^+) &=(n_{12}^++\cdots+n_{1,N}^++2n_1^+,\cdots,n_{1,N}^++\cdots+n_{N-1,N}^++2n_N^+).
    \end{split}
\end{align*}
Following same procedure we did in the $SO(2N)$ case, we get a combinatorial like formula for deformed Schur index of $Sp(2N)$ $\mathcal{N}=4$ SYM theory:
\begin{align}
\begin{split}
&I_{Sp(2N)} (u,q,t)  = \mathcal{X}_N  \sum_{\ell(\lambda) \leq N} u^{|\lambda|} b_\lambda \\
&\sum_{{n_{12}^+\geq\cdots \geq n_{N-1,N}^+\geq n_1^+\geq \cdots\geq n_N^+}}^{\infty}\ \ \sum_{{n_{12}^-\geq\cdots \geq n_{N-1,N}^-\geq n_1^-\geq \cdots\geq n_N^-}}^{\infty}c_{n_{12}^+,\cdots ,n_{N-1,N}^+,n_1^+,\cdots, n_N^+} c_{{n_{12}^-,\cdots ,n_{N-1,N}^-,n_1^-,\cdots, n_N^-}}\\[10pt]
&\quad\quad\quad\quad\quad\quad\ \times\sideset{}{^{\{n^+\}}}\sum_{\mu}\sideset{}{^{\{n^-\}}}\sum_{\nu}\left\langle n_{\mu}m_{\mu}(\mathbf{x})P_{\lambda}(\mathbf{x}), n_{\nu}m_{\nu}(\mathbf{x})P_{\lambda}(\mathbf{x})\right \rangle.
\end{split}
\end{align}
 In general:
\begin{align*}
&I_{Sp(2N)} (u,q,t)\\
&=\mathcal{X}_N \Bigg[\mathcal{N}_{[0],N}+ub_{[1]}\mathcal{N}_{[1],N}+\Bigg(u^2b_{[1,1]}+c_{1}^2\Bigg(1+\frac{-qt+q-t+1}{qt-1}\Bigg)^2\Bigg)\mathcal{N}_{[1,1],N}&\\
&\quad\quad+\Bigg(u^2b_{[2]}+c_{1}^2\Bigg)\mathcal{N}_{[2],N}+\Bigg(ub_{[1]}c_{1,0,\cdots,0}^2\Bigg(1+\frac{q^2 - q t + q - t}{q^2 t - 1}\Bigg)^2+u^3b_{[2,1]}\Bigg)\mathcal{N}_{[2,1],N} &\\
&\quad\quad+\Bigg(u^3b_{[3]}+ub_{[1]}c_{1}^2\Bigg)\mathcal{N}_{[3],N}+\Bigg(u^3b_{[1,1,1]}+uc_{1,0,\cdots,0}^2b_{[1]}\Bigg( W+X \Bigg)^2\Bigg)\mathcal{N}_{[1,1,1],N} &\\
&\quad\quad+\Bigg(u^4b_{[1,1,1,1]}+u^2b_{[1,1]}c_{1}^2\Bigg(M+O\Bigg)^2+\Bigg(c_{1,1}\Bigg(3+2L+A+D\Bigg)+c_{2}\Bigg(D+F\Bigg) \Bigg)^2\Bigg)\mathcal{N}_{[1,1,1,1],N}&\\
&\quad\quad+\Bigg(u^4b_{[2,1,1]}+u^2c_{1}^2\Bigg(b_{[1,1]}\Bigg(N+Q\Bigg)^2+b_{[2]}\Bigg(S+T\Bigg)^2\Bigg)&\\
&\quad\quad\quad\quad\quad\quad\quad\quad\quad\quad\quad\quad+\Bigg(c_{1,1}\Bigg(2+B+E\Bigg)+c_{2}\Bigg(E+G\Bigg) \Bigg)^2\Bigg)\mathcal{N}_{[2,1,1],N}&\\
&\quad\quad +\Bigg(u^4b_{[3,1]}+u^2c_{1}^2\Bigg(b_{[1,1]}\Bigg(1+R\Bigg)^2+b_{[2]}U^2\Bigg(S+T\Bigg)^2\Bigg)\\
&\quad\quad\quad\quad\quad\quad\quad\quad\quad\quad\quad\quad\quad\quad+\Bigg(c_{1,1}\Bigg(C+1\Bigg)+c_{2}\Bigg(I+1\Bigg)\Bigg)^2\Bigg)\mathcal{N}_{[2,2],N} &\\
&\quad\quad +\Bigg(u^4b_{[3,1]}+u^2c_{1}^2\Bigg(b_{[1,1]}+b_{[2]}\Bigg(1+K\Bigg)^2\Bigg)+\Bigg(c_{1,1}+c_{2}H\Bigg)^2\Bigg)\mathcal{N}_{[3,1],N}&\\
&\quad\quad+\Bigg(u^4b_{[4]}+u^2c_{1}^2b_{[2]}+c_{2}^2\Bigg)\mathcal{N}_{[4],N}+\mathcal{O}\Bigg(\mathcal{N}_{|\lambda|>4,N}\Bigg) \Bigg],
\end{align*}
where
\begin{align*}
    &A=\frac{2q^2t^5-q^2t^4+qt^5-2q^2t^3+qt^4+t^5-2qt^3-2qt^2+q^2+qt-2t^2+q-t+2}{q^2t^5-qt^3-qt^2+1},\\[1ex]
    &B=\frac{-q^3t^3+q^3t+t^2-1}{q^3t^3-q^2t^2-qt+1},\quad C=\frac{-qt+q-t+1}{qt-1},\\[1ex]
    &D=\frac{q^2t^5-q^2t^4+qt^5-q^2t^2-qt^3+t^4+q^2t-qt^2-t^3+q-t+1}{q^2t^5-qt^3-qt^2+1},\
    E=\frac{-qt+q-t+1}{qt-1},\\[1ex]
    &F=\scalebox{0.95}{$\frac{-q^3t^6+q^3t^5-q^2t^6+q^3t^4+q^2t^5-qt^6+q^2t^4+qt^5-t^6-q^3t^2+qt^4+t^5-q^3t-q^2t^2+t^4+q^3-q^2t-qt^2+q^2-qt-t^2+q-t+1}{q^3t^6-q^2t^5-q^2t^4-q^2t^3+qt^3+qt^2+qt-1}$},\\[1ex]
    &G=\frac{q^3t^3-q^3t^2+q^2t^3-q^3t-q^2t^2+qt^3+q^3-q^2t-qt^2+t^3+q^2-qt-t^2+q-t+1}{q^3t^3-q^2t^2-qt+1},\\[1ex]
    &H=\frac{-q^3t+q^2t^2+q^3-q^2t-qt+t^2+q-t}{q^3t^2-q^2t-qt+1},\ I=\frac{-q^3t+q^2t^2+q^3-q^2t-qt+t^2+q-t}{q^3t^2-q^2t-qt+1},\\[1ex]
    & K=\frac{q^3 t - q^2 t^2 - q^3 + 2 q^2 t - q t^2 - q^2 + q t}{q^4 t^2 - q^3 t - q t + 1},\quad L=\frac{-3qt^3+qt^2-t^3+qt-t^2+q-t+3}{qt^3-1},\\[1ex]
    &M=\scalebox{1.3}{$\frac{q^2t^5+q^2t^4-qt^5+2q^2t^3-2qt^4+q^2t^2-3qt^3+t^4+q^2t-3qt^2+t^3-2qt+2t^2-q+t+1}{q^2t^5-qt^3-qt^2+1}$},\\[1ex]
    &N=\frac{qt+q-t-1}{qt-1},\quad O=\frac{-q^2t^5-q^2t^3+t^5+q^2t^2+t^3+q^2-t^2-1}{q^2t^5-qt^3-qt^2+1}\\
    &Q=\frac{q^3t-q^2t^2-qt+t^2}{q^3t^3-q^2t^2-qt+1},\quad R=\frac{-qt+q-t+1}{qt-1},\quad S=\frac{q^3t^3-qt^3-q^2+1}{q^3t^3-q^2t^2-qt+1},\\[1ex]
    &T=\scalebox{1.3}{$\frac{-q^4t^4+q^4t^3-q^3t^4+q^3t^3+q^2t^4-q^2t^3+qt^4+q^3t-qt^3-q^3+q^2t-q^2-qt+q-t+1}{q^4t^4-2q^3t^3+2qt-1}$},\\[1ex]
    &U=\scalebox{1.3}{$\frac{q^4t^3+q^4t^2-q^3t^3-q^3t^2-q^2t^3-q^3t-q^2t^2+qt^3-q^3+q^2t+qt^2+q^2+qt+q-t-1}{q^4t^3-2q^3t^2-q^2t^2+q^2t+2qt-1}$},\\[1ex]
    &W=\frac{-qt^2-qt+t^2-q+t+1}{-qt^2+1},\quad X=\frac{q^2t^3-t^3-q^2+1}{-q^2t^3+qt^2+qt-1},\quad Y=\frac{q^2 - q t + q - t}{q^2 t - 1}.
\end{align*}
If we fix a number $N$, the normalization factor $\mathcal{N}_{\ell(\lambda)>N,N}$ will vanish, this is the mechanism by which $N$ enters our computation: it imposes a length constraint on all partitions involved in the calculation in the deformed Schur indices.

For $SO(2N+1)$, the case is almost same as $Sp(2N)$ ,the only difference is:
\begin{align*}
       &A'(n_{12}^+,n_{13}^+,\cdots ,n_{N-1,N}^+,n_1^+,n_2^+,\cdots, n_N^+)\to A''(n_{12}^+,n_{13}^+,\cdots ,n_{N-1,N}^+,n_1^+,n_2^+,\cdots, n_N^+),\\
       &A'(n_{12}^+,\cdots ,n_{N-1,N}^+,n_1^+,\cdots, n_N^+) =(n_{12}^++\cdots+n_{1,N}^++2n_1,\cdots,n_{1,N}^++\cdots+n_{N-1,N}^++2n_N),\\
       &A''(n_{12}^+,\cdots ,n_{N-1,N}^+,n_1^+,\cdots, n_N^+)=(n_{12}^++\cdots+n_{1,N}^++n_1,\cdots,n_{1,N}^++\cdots+n_{N-1,N}^++n_N).  
\end{align*}
So the result can be gotten directly:
\begin{align}
\begin{split}
&I_{SO(2N+1)} (u,q,t)  = \mathcal{X}_N  \sum_{\ell(\lambda) \leq N} u^{|\lambda|} b_\lambda \\
&\sum_{{n_{12}^+\geq\cdots \geq n_{N-1,N}^+\geq n_1^+\geq \cdots\geq n_N^+}}^{\infty}\ \ \sum_{{n_{12}^-\geq\cdots \geq n_{N-1,N}^-\geq n_1^-\geq \cdots\geq n_N^-}}^{\infty}c_{n_{12}^+,\cdots ,n_{N-1,N}^+,n_1^+,\cdots, n_N^+} c_{{n_{12}^-,\cdots ,n_{N-1,N}^-,n_1^-,\cdots, n_N^-}}\\[10pt]
&\quad\quad\quad\quad\quad\quad\quad\ \times\sideset{}{^{\{n^+\}}}\sum_{\mu}\sideset{}{^{\{n^-\}}}\sum_{\nu}\left\langle n_{\mu}m_{\mu}(\mathbf{x})P_{\lambda}(\mathbf{x}), n_{\nu}m_{\nu}(\mathbf{x})P_{\lambda}(\mathbf{x})\right \rangle.
\end{split}
\end{align}
In general:
\begin{align*}
&I_{SO(2N+1)} (u,q,t)\\
&=\mathcal{X}_N \Bigg[\mathcal{N}_{[0],2}+\Bigg(ub_{[1]}+c_{1}^2\Bigg)\mathcal{N}_{[1],N}+\Bigg(u^2b_{[2]}+c_{2}^2+ub_{[1]}c_{1}^2\Bigg)\mathcal{N}_{[2],N}&\\
&\quad\quad+\Bigg(u^2b_{[1,1]}+\Bigg(c_{1}+c_{1,1}+lc_{2}\Bigg)^2+um^2b_{[1]}c_{1}^2\Bigg)\mathcal{N}_{[1,1],N}\\
&\quad\quad+\Bigg(u^3b_{[2,1]}+\Bigg(c_{1,1}+c_{2,1}+kc_{3}\Bigg)^2+ub_{[1]}\Bigg(c_{1}+c_{1,1}+Yc_{2}\Bigg)^2\\
&\quad\quad\quad\quad\quad\quad+\Bigg(u^2b_{[1,1]}+u^2b_{[2]}p^2\Bigg)c_{1}^2\Bigg)\mathcal{N}_{[2,1],N} &\\
&\quad\quad+\Bigg(u^3b_{[3]}+c_{3,0,\cdots,0}^2+ub_{[1]}c_{2,0,\cdots,0}^2+u^2b_{[2]}c_{1,0,\cdots,0}^2\Bigg)\mathcal{N}_{[3],N}\\
&\quad\quad+\Bigg(u^3b_{[1,1,1]}+\Bigg(3c_{1,1}+c_{1,1,1}+i\Bigg(c_{1,1}+c_{2,1}\Bigg)+jc_{3}\Bigg)^2\\
&\quad\quad\quad\quad+ub_{[1]}\Bigg(\Bigg(c_{1}+c_{1,1}\Bigg)W+Xc_{2}\Bigg)^2+u^2b_{[1,1]}n^2c_{1}^2\Bigg)\mathcal{N}_{[1,1,1],N}+
\mathcal{O}\Bigg( \mathcal{N}_{|\lambda|>3,N} \Bigg) \Bigg],
\end{align*}
where the uppercase letters retain their previous definitions in $Sp(2N)$ case, while lowercase letters are defined as follows:
\begin{align*}
&l=\frac{-qt+q-t+1}{qt-1},\quad m=\frac{qt+q-t-1}{qt-1},\quad n=\frac{-qt^2-qt+t^2-q+t+1}{-qt^2+1},\\
&i=\frac{2qt^2-qt+t^2-q+t-2}{-qt^2+1},\ k=\frac{q^2t-q^2+qt-q+t-1}{-q^2t+1},\\
&j=\frac{-q^2t^3+q^2t^2-qt^3+q^2t+qt^2-t^3-q^2+qt+t^2-q+t-1}{-q^2t^3+qt^2+qt-1},\\
&p=\frac{-q^3t^2+qt^2+q^2-1}{-q^3t^2+q^2t+qt-1}.
\end{align*}
\\

\section{Expansion in powers of parameter $u$}
\label{sec4}
In this section, we start with the expression of the $BCD_N$ type deformed Schur indices (\ref{KoornwidnerSp(2N)aftertheorem2}), (\ref{KoornwidnerSO(2N)aftertheorem2}), (\ref{KoornwidnerSO(2N+1)aftertheorem2}), (\ref{KoornwidnerSO(2N+1)aftertheorem2andspecexpans}) together with (\ref{KoornwidnerSO(2N+1)âaftertheorem2}). In these cases, we expand the Koornwinder weight function excludes terms already accounted for in the $U(N)$ case (\ref{weightfunction}), then restructure it into monomial symmetric functions, and define: 
\begin{equation*}
    d_{n}(q,t):=\frac{(t^{-1};q)_{n}}{(q;q)_{n}}t^{n},\quad\quad d_{n_1,n_2,\cdots ,n_N}(q,t):=d_{n_1}(q,t)d_{n_2}(q,t)\cdots d_{n_N}(q,t).
\end{equation*}
Following same procedure introduced in Section \ref{sec3}, without repeating the derivation, we will present the final results directly:
\begin{align}
\begin{split}
I_{SO(2N)} (u,q,t)  = &\mathcal{X}_N \sum_{\ell(\lambda) \leq N}\sum_{\lambda}\sum_{\text{$m_i(\mu)$,$m_i(\nu)$is even}}u^{|\lambda|+\frac{|\mu|+|\nu|}{2}}b_{\lambda}(q,t)b^{\mathrm{ol}}_{\mu}(q,t)b^{\mathrm{ol}}_{\nu}(q,t)\\
&\times\sum_{{n_{12}^+\geq n_{13}^+\geq\cdots \geq n_{N-1,N}^+}}^{\infty}\sum_{{n_{12}^-\geq n_{13}^-\geq\cdots \geq n_{N-1,N}^-}}^{\infty}d_{n_{12}^+,n_{13}^+,\cdots ,n_{N-1,N}^+} d_{{n_{12}^-,n_{13}^-,\cdots ,n_{N-1,N}^-}}\\
& \times\sideset{}{^{\{n^+\}}}\sum_{\rho}\sideset{}{^{\{n^-\}}}\sum_{\gamma}\left\langle n_{\rho}m_{\rho}(\mathbf{x})P_{\mu}(\mathbf{x})P_{\lambda}(\mathbf{x}), n_{\gamma}m_{\gamma}(\mathbf{x})P_{\nu}(\mathbf{x})P_{\lambda}(\mathbf{x})\right \rangle,
\end{split}
\end{align}
\begin{align}
\begin{split}
&I_{Sp(2N)} (u,q,t)  = \mathcal{X}_N \sum_{\ell(\lambda) \leq N}\sum_{\text{$|\mu|$,$|\nu|$ is even}}u^{|\lambda|+\frac{|\mu|+|\nu|}{2}}b_{\lambda}(q,t)b^{\mathrm{e}}_{\mu}(q,t)b^{\mathrm{e}}_{\nu}(q,t)\\
&\sum_{{n_{12}^+\geq\cdots \geq n_{N-1,N}^+\geq n_1^+\geq \cdots\geq n_N^+}}^{\infty}
\sum_{{n_{12}^-\geq \cdots \geq n_{N-1,N}^-\geq n_1^-\geq \cdots\geq n_N^-}}^{\infty}d_{n_{12}^+,\cdots ,n_{N-1,N}^+,n_1^+,\cdots, n_N^+} d_{{n_{12}^-,\cdots ,n_{N-1,N}^-,n_1^-,\cdots, n_N^-}}\\[10pt]
&\quad\quad\quad\quad\quad\quad\ \times\sideset{}{^{\{n^+\}}}\sum_{\rho}\sideset{}{^{\{n^-\}}}\sum_{\gamma}\left\langle n_{\rho}m_{\rho}(\mathbf{x})P_{\mu}(\mathbf{x})P_{\lambda}(\mathbf{x}), n_{\gamma}m_{\gamma}(\mathbf{x})P_{\nu}(\mathbf{x})P_{\lambda}(\mathbf{x})\right \rangle,
\end{split}
\end{align}
\begin{align}
\begin{split}
&I'_{SO(2N+1)} (u,q,t)  = \mathcal{X}_N \sum_{\ell(\lambda) \leq N}\sum_{\mu,\nu}u^{|\lambda|+\frac{|\mu|+|\nu|}{2}}b_{\lambda}(q,t)b^{\mathrm{el}}_{\mu}(q,t)b^{\mathrm{el}}_{\nu}(q,t)\\
&\sum_{{n_{12}^+\geq\cdots \geq n_{N-1,N}^+\geq n_1^+\geq \cdots\geq n_N^+}}^{\infty}\ \ \sum_{{n_{12}^-\geq \cdots \geq n_{N-1,N}^-\geq n_1^-\geq \cdots\geq n_N^-}}^{\infty}d_{n_{12}^+,\cdots ,n_{N-1,N}^+,n_1^+,\cdots, n_N^+} d_{{n_{12}^-,\cdots ,n_{N-1,N}^-,n_1^-,\cdots, n_N^-}}\\[10pt]
&\quad\quad\quad\quad\quad\quad\quad\ \times\sideset{}{^{\{n^+\}}}\sum_{\rho}\sideset{}{^{\{n^-\}}}\sum_{\gamma}\left\langle n_{\rho}m_{\rho}(\mathbf{x})P_{\mu}(\mathbf{x})P_{\lambda}(\mathbf{x}), n_{\gamma}m_{\gamma}(\mathbf{x})P_{\nu}(\mathbf{x})P_{\lambda}(\mathbf{x})\right \rangle,
\end{split}
\end{align}
\begin{align}
    \begin{split}
    &I_{SO(2N+1)} (u,q,t)  = \mathcal{X}_N \sum_{\lambda}\sum_{\mu,\nu}u^{|\lambda|+\frac{|\mu|+|\nu|+odd(\mu')+odd(\nu')}{2}}b_{\lambda}(q,t)b^{\mathrm{el}}_{\mu}(q,t)b^{\mathrm{el}}_{\nu}(q,t)\\
&\sum_{{n_{12}^+\geq\cdots \geq n_{N-1,N}^+\geq n_1^+\geq \cdots\geq n_N^+}}^{\infty}\ \ \sum_{{n_{12}^-\geq \cdots \geq n_{N-1,N}^-\geq n_1^-\geq \cdots\geq n_N^-}}^{\infty}d_{n_{12}^+,\cdots ,n_{N-1,N}^+,n_1^+,\cdots, n_N^+} d_{{n_{12}^-,\cdots ,n_{N-1,N}^-,n_1^-,\cdots, n_N^-}}\\[10pt]
&\quad\quad\quad\quad\quad\quad\quad\ \times\sideset{}{^{\{n^+\}}}\sum_{\rho}\sideset{}{^{\{n^-\}}}\sum_{\gamma}\left\langle n_{\rho}m_{\rho}(\mathbf{x})P_{\mu}(\mathbf{x})P_{\lambda}(\mathbf{x}), n_{\gamma}m_{\gamma}(\mathbf{x})P_{\nu}(\mathbf{x})P_{\lambda}(\mathbf{x})\right \rangle,
    \end{split}
\end{align}
or
\begin{align}
    \begin{split}
    I_{SO(2N+1)} (u,q,t)  &= \mathcal{X}_N \sum_{l=0}^\infty\sum_{\lambda}\sum_{\text{$m_i(\mu)$,$m_i(\nu)$is even}}u^{|\lambda|+l+\frac{|\mu|+|\nu|}{2}}b_{\lambda}(q,t)b^{\mathrm{ol}}_{\mu}(q,t)b^{\mathrm{ol}}_{\nu}(q,t)w_l(q,t)\\
&\times\sum_{{n_{12}^+\geq n_{13}^+\geq\cdots \geq n_{N-1,N}^+}}^{\infty}\sum_{{n_{12}^-\geq n_{13}^-\geq\cdots \geq n_{N-1,N}^-}}^{\infty}d_{n_{12}^+,n_{13}^+,\cdots ,n_{N-1,N}^+} d_{{n_{12}^-,n_{13}^-,\cdots ,n_{N-1,N}^-}}\\
& \times\sideset{}{^{\{n^+\}}}\sum_{\rho}\sideset{}{^{\{n^-\}}}\sum_{\gamma}\left\langle n_{\rho}m_{\rho}(\mathbf{x})P_{\mu}(\mathbf{x})P_{\lambda}(\mathbf{x})P_{[l]}(\mathbf{x}), n_{\gamma}m_{\gamma}(\mathbf{x})P_{\nu}(\mathbf{x})P_{\lambda}(\mathbf{x})P_{[l]}(\mathbf{x})\right \rangle.
    \end{split}
\end{align}

By expanding these formulas with normalization factors $\mathcal{N}_{\lambda,N}$, we verify their equivalence to the results obtained in Section \ref{sec3}. This equivalence reveals a set of non-trivial identities relating the specialized $q$-Pochhammer symbols $c_{n_1,\dots,n_N}(u;q,t)$ and $d_{n_1,\dots,n_N}(q,t)$ to the combinatorial quantities $b_{\lambda}(q,t)$, $b_{\lambda}^{\text{ol}}(q,t)$, $b_{\lambda}^{\text{oa}}(q,t)$, $b_{\lambda}^{\text{el}}(q,t)$ and $q,t$-deformed Littlewood-Richardson coefficients.

\section{Some further limits}\label{sec5}
Further limits based on the $p = v = 0$ case will also be interesting. In $u\to 0$ limit, in our formalism, only trivial Macdonald polynomial $P_{\emptyset}(\mathbf{x};q,t)=1$ contributes, then the integrals (\ref{Cdsi}), (\ref{Bdsi}), (\ref{Ddsi}) reduce to the integrals called the multivariable extension of the Askey-Wilson integrals (or particular q-Selberg integrals serving as the orthogonality measure for Koornwinder polynomials), and they have been evaluated exactly in \cite{MR1001607} \cite{MR1199128}:
\begin{align*}
I_{Sp(2N)}^{p=v=u=0} &= I_{SO(2N+1)}^{p=v=u=0} = \prod_{j=0}^{N-1} \frac{(q t^{2j+1};q)_\infty}{(t^{2j+2};q)_\infty},\\
I_{SO(2N)}^{p=v=u=0} &= \frac{(qt^{N-1};q)_\infty}{(t^N;q)_\infty}\prod_{j=0}^{N-2}\frac{(qt^{2j+1};q)_\infty}{(t^{2j+2};q)_\infty}.
\end{align*}
S-duality for $\mathcal{N} = 4$ $SO(2N + 1)/Sp(2N)$ SYM\cite{gadde2010s} is manifest in this limit. Due to the symmetry under the exchange of parameters $t$ and $u$, we can immediately obtain analogous results in  $t \to 0$ limit through substitution $t \to u $ in these results.

Another important limiting case occurs when $q=t$. In this limit, Macdonald polynomials will reduce to the ordinary Schur function  $s_{\lambda}(\mathbf{x})$. In the  $BCD_N$ case, this correspondence generalizes to Koornwinder polynomials, which similarly reduce to the symplectic, orthogonal and odd-orthogonal Schur functions in our speializations of parameters respectively \cite{MR4191808}:
\begin{align*}
\mathrm{sp}_{\lambda}(x_1,\ldots,x_n) &= K_\lambda\big(x_1,\ldots,x_n;q,q;q^{1/2},-q^{1/2},q,-q\big), \\
\mathrm{o}_{\lambda}(x_1,\ldots,x_n) &= K_\lambda\big(x_1,\ldots,x_n;q,q;1,-1,q^{1/2},-q^{1/2}\big), \\
\mathrm{so}_{\lambda}(x_1,\ldots,x_n) &= K_\lambda\big(x_1,\ldots,x_n;q,q;-1,-q^{1/2},q^{1/2},q\big).
\end{align*}
All theorems established in Section \ref{sec2} continue to hold when $q = t$, and in this limit, the specializations $(+\sqrt{q},-\sqrt{q},+\sqrt{qt},-\sqrt{qt})$, $(+\sqrt{t},-\sqrt{t},+\sqrt{qt},-\sqrt{qt})$ coincide, so the first line in Theorem.\ref{thm2} can be used for the $Sp(2N)$ case. Consequently, we can directly write down the indices in this limiting case as follows:
\begin{align}
I_{Sp(2N)}^{p=v=0,q=t} (u,q)  &=
\begin{aligned}[t]
    &\mathcal{X}_N \sum_{\lambda}\sum_{\text{$\mu$,$\nu$ is even}}u^{|\lambda|+\frac{|\mu|+|\nu|}{2}}b_{\lambda}(q,q)b^{\mathrm{oa}}_{\mu}(q,q)b^{\mathrm{oa}}_{\nu}(q,q)\\
&\times \oint_{\mathbb{T}^N} \ \Delta_{C_N} s_{\lambda}(\mathbf{x}) s_{\mu}(\mathbf{x})s_{\lambda}(\mathbf{x}^{-1})s_{\nu}(\mathbf{x}^{-1})\prod_{j=1}^N\frac{dx_j}{2\pi ix_j},
\label{sc$Sp(2N)$}
\end{aligned}\\[2ex]
\label{sc$SO(2N+1)'$}
I_{SO(2N+1)}^{'s=v=0,q=t} (u,q)  &= 
\begin{aligned}[t]
&\mathcal{X}_N \sum_{\lambda}\sum_{\mu,\nu}u^{|\lambda|+\frac{|\mu|+|\nu|}{2}}b_{\lambda}(q,q)b^{\mathrm{el}}_{\mu}(q,q)b^{\mathrm{el}}_{\nu}(q,q)\\
&\times \oint_{\mathbb{T}^N} \ \Delta_{B_N}s_{\lambda}(\mathbf{x}) s_{\mu}(\mathbf{x})s_{\lambda}(\mathbf{x}^{-1})s_{\nu}(\mathbf{x}^{-1})\prod_{j=1}^N\frac{dx_j}{2\pi ix_j},
\end{aligned}\\[2ex]
\label{sc$SO(2N)$}
I_{SO(2N)}^{s=v=0,q=t} (u,q)  &=
\begin{aligned}[t]
&\mathcal{X}_N \sum_{\lambda}\sum_{\substack{\text{$m_i(\mu),m_i(\nu)$} \\[3pt] \text{is even}}}u^{|\lambda|+\frac{|\mu|+|\nu|}{2}}b_{\lambda}(q,q)b^{\mathrm{ol}}_{\mu}(q,q)b^{\mathrm{ol}}_{\nu}(q,q)\\
&\times \oint_{\mathbb{T}^N} \ \Delta_{D_N}s_{\lambda}(\mathbf{x}) s_{\mu}(\mathbf{x})s_{\lambda}(\mathbf{x}^{-1})s_{\nu}(\mathbf{x}^{-1})\prod_{j=1}^N\frac{dx_j}{2\pi ix_j},
\end{aligned}\\[2ex]
\label{sc$SO(2N+1)$1}
I_{SO(2N+1)}^{s=v=0,q=t} (u,q)  &= 
\begin{aligned}[t]
&\mathcal{X}_N \sum_{\lambda}\sum_{\mu,\nu}u^{|\lambda|+\frac{|\mu|+|\nu|+odd(\mu')+odd(\nu')}{2}}b_{\lambda}(q,q)b^{\mathrm{el}}_{\mu}(q,q)b^{\mathrm{el}}_{\nu}(q,q)\\
&\times \oint_{\mathbb{T}^N} \ \Delta_{B_N}s_{\lambda}(\mathbf{x}) s_{\mu}(\mathbf{x})s_{\lambda}(\mathbf{x}^{-1})s_{\nu}(\mathbf{x}^{-1})\prod_{j=1}^N\frac{dx_j}{2\pi ix_j},
\end{aligned}\\[2ex]
\label{sc$SO(2N+1)$2}
I_{SO(2N+1)}^{s=v=0,q=t} (u,q)  &= 
\begin{aligned}[t]
&\mathcal{X}_N \sum_{\lambda}\sum_l\sum_{\substack{\text{$m_i(\mu),m_i(\nu)$} \\[3pt] \text{is even}}}u^{|\lambda|+l+\frac{|\mu|+|\nu|}{2}}b_{\lambda}(q,q)b^{\mathrm{el}}_{\mu}(q,q)b^{\mathrm{el}}_{\nu}(q,q)w_l(q,q)\\
&\times \oint_{\mathbb{T}^N} \ \Delta_{B_N}s_{\lambda}(\mathbf{x}) s_{\mu}(\mathbf{x})s_{[l]}(\mathbf{x})s_{\lambda}(\mathbf{x}^{-1})s_{\nu}(\mathbf{x}^{-1})s_{[l]}(\mathbf{x}^{-1})\prod_{j=1}^N\frac{dx_j}{2\pi ix_j},
\end{aligned}
\end{align}
where the $\Delta_{B_N},\Delta_{C_N},\Delta_{D_N}$ are reduced koornwinder weight functions in $q=t$ limit.
And $b_{\lambda}(q,q)=b^{\mathrm{ol}}_{\lambda}(q,q)=b^{\mathrm{oa}}_{\lambda}(q,q)=b^{\mathrm{el}}_{\lambda}(q,q)=w_l(q,q)=1$ for arbitrary $\lambda$ and $l$. (\ref{sc$SO(2N+1)$1}) and (\ref{sc$SO(2N+1)$2})  yield identical results when expanded according to the final line of Theorem \ref{thm2} and (\ref{expansion1}), respectively.  To calculate the indices in the limit $q=t$, we need to expand the part of weight function that differs from the A-type weight function in the Schur function basis. For example, in the $C_N$ case, we expand $\prod_{i=1}^N(1-x_i^2)\prod_{1\leq i<j\leq N}(1-x_ix_j)$ in the Schur basis. Unlike the deformed Schur index case where expansions involve infinite monomial symmetric functions, this expansion can be performed recursively in $N$ and involves only a finite number of Schur functions for finite $N$:
\begin{align*}
    \prod_{i=1}^N(1-x_i^2)\prod_{1\leq i<j\leq N}(1-x_ix_j)&=\sum_{\lambda \subset S^C_{N}}f^C_{\lambda}s_{\lambda}:=\\
    &(-1)^{\frac{N(N+1)}{2}}\left(\sum_{\lambda \subset S^C_{N-1}}f^C_{\lambda}s_{[(N+1)^N-\lambda]}\right) + \sum_{\lambda \subset S^C_{N-1}}f^C_{\lambda}s_{\lambda},
\end{align*}
where $f_{\lambda}=\{+1,-1\}$,  $S^C_{N}$ denote the collection of Young diagrams associated with the Schur functions emerging in the expansion of the product $\prod_{i=1}^N(1-x_i^2)\prod_{1\leq i<j\leq N}(1-x_ix_j)$, and
\begin{align*}
    \prod_{i=1}^{N-1}(1-x_i^2)\prod_{1\leq i<j\leq N-1}(1-x_ix_j)= (-1)^{\frac{N(N+1)}{2}}\sum_{\lambda \subset S^C_{N-1}}f^C_{\lambda}s_{\lambda}.
\end{align*}
This recursion procedure start from:
\begin{align*}
    1&=s_{[\emptyset]},\\
    1-x_1^2&=-s_{[2]}(x_1)+s_{[\emptyset]}(x_1),\\
    (1-x_1^2)(1-x_2^2)( 1-x_1x_2)&=-s_{[3,3]}(x_1,x_2)+s_{[3,1]}(x_1,x_2)-s_{[2]}(x_1,x_2)+s_{[\emptyset]}(x_1,x_2),
\end{align*}
$S^C_{0}=\{[\emptyset]\}$, $S^C_{1}=\{[\emptyset],[2]\}$ and let $|S_N^C|$ denote the numbers of young diagrams in $S_N^C$ . $S_N^C$ can be defined recursively by $S_N^C=S_{N-1}^C\cup \{(N+1)^N-\lambda|\lambda\in S_{N-1}^C\}$, so $|S_N^C|=2^N$. This recursive relation similarly holds for $BD_N$-type cases, in $B_N$ case:
\begin{align*}
    \prod_{i=1}^N(1-x_i)\prod_{1\leq i<j\leq N}(1-x_ix_j)&=\sum_{\lambda \subset S^B_{N}}f^B_{\lambda}s_{\lambda}:=\\
    &(-1)^{\frac{N(N+1)}{2}}\left(\sum_{\lambda \subset S^B_{N-1}}f^B_{[N^N-\lambda]}s_{\lambda}\right) + \sum_{\lambda \subset S^B_{N-1}}f^B_{\lambda}s_{\lambda},
\end{align*}
where
\begin{align*}
    \prod_{i=1}^{N-1}(1-x_i^2)\prod_{1\leq i<j\leq N-1}(1-x_ix_j)= (-1)^{\frac{N(N+1)}{2}}\sum_{\lambda \subset S^B_{N-1}}f^B_{\lambda}s_{\lambda}.
\end{align*}
And this recursion procedure start from:
\begin{align*}
1&=s_{[\emptyset]},\\
    1-x_1&=-s_{[1]}(x_1)+s_{[\emptyset]}(x_1),\\
    (1-x_1)(1-x_2)( 1-x_1x_2)&=-s_{[2,2]}(x_1,x_2)+s_{[2,1]}(x_1,x_2)-s_{[1]}(x_1,x_2)+s_{[\emptyset]}(x_1,x_2),
\end{align*}
$S^B_{0}=\{[\emptyset]\}$, $S^B_{1}=\{[\emptyset],[1]\}$and $|S_N^B|=2^N$. In $D_N$ case:
\begin{align*}
    \prod_{1\leq i<j\leq N}(1-x_ix_j)&=\sum_{\lambda \subset S^D_{N}}f^D_{\lambda}s_{\lambda}:=(-1)^{\frac{N(N-1)}{2}}\left(\sum_{\lambda \subset S^D_{N-1}}f^D_{\lambda}s_{[(N-1)^N-\lambda]}\right) + \sum_{\lambda \subset S^D_{N-1}}f^D_{\lambda}s_{\lambda},
\end{align*}
where
\begin{align*}
    \prod_{1\leq i<j\leq N-1}(1-x_ix_j)= (-1)^{\frac{N(N-1)}{2}}\sum_{\lambda \subset S^D_{N-1}}s_{\lambda}.
\end{align*}
And this recursion procedure start from:
\begin{align*}
    1&=s_{[\emptyset]},\\
    1-x_1x_2&=-s_{[1,1]}(x_1,x_2)+s_{[\emptyset]}(x_1,x_2),\\
    (1-x_1x_2)(1-x_1x_3)(1-x_2x_3)&=-s_{[2,2,2]}(x_1,x_2,x_3)+s_{[2,1,1]}(x_1,x_2,x_3)\\
    &\quad\ -s_{[1,1]}(x_1,x_2,x_3)+s_{[\emptyset]}(x_1,x_2,x_3),
\end{align*}
$S^D_{1}=\{[\emptyset]\}$, $S^D_{2}=\{[\emptyset],[1,1]\}$ and $|S_N^D|=2^{N-1}$. Using these identities, we can rewrite the indices as:
\begin{align}
I_{Sp(2N)}^{p=v=0,q=t} (u)  &=
\begin{aligned}[t]
    &\mathcal{X}_N \sum_{\rho\subset S_N^C}\sum_{\gamma\subset S_N^C}\sum_{\lambda}\sum_{\text{$\mu$,$\nu$ is even}}u^{|\lambda|+\frac{|\mu|+|\nu|}{2}}f^C_{\rho}f^C_{\gamma}\\
&\times \oint_{\mathbb{T}^N} \ \Delta_{A_{N-1}} s_{\lambda}(\mathbf{x}) s_{\mu}(\mathbf{x})s_{\rho}(\mathbf{x})s_{\lambda}(\mathbf{x}^{-1})s_{\nu}(\mathbf{x}^{-1})s_{\gamma}(\mathbf{x}^{-1})\prod_{j=1}^N\frac{dx_j}{2\pi ix_j},
\label{sc$Sp(2N)$2}
\end{aligned}\\[2ex]
\label{sc$SO(2N+1)'$2}
I_{SO(2N+1)}^{'s=v=0,q=t} (u)  &= 
\begin{aligned}[t]
&\mathcal{X}_N \sum_{\rho\subset S_N^B}\sum_{\gamma\subset S_N^B}\sum_{\lambda}\sum_{\mu,\nu}u^{|\lambda|+\frac{|\mu|+|\nu|}{2}}f^B_{\rho}f^B_{\gamma}\\
&\times \oint_{\mathbb{T}^N} \ \Delta_{A_{N-1}} s_{\lambda}(\mathbf{x}) s_{\mu}(\mathbf{x})s_{\rho}(\mathbf{x})s_{\lambda}(\mathbf{x}^{-1})s_{\nu}(\mathbf{x}^{-1})s_{\gamma}(\mathbf{x}^{-1})\prod_{j=1}^N\frac{dx_j}{2\pi ix_j},
\end{aligned}\\[2ex]
\label{sc$SO(2N)$2}
I_{SO(2N)}^{s=v=0,q=t} (u)  &=
\begin{aligned}[t]
&\mathcal{X}_N \sum_{\rho\subset S_N^D}\sum_{\gamma\subset S_N^D}\sum_{\lambda}\sum_{\substack{\text{$m_i(\mu),m_i(\nu)$} \\[3pt] \text{is even}}}u^{|\lambda|+\frac{|\mu|+|\nu|}{2}}f^D_{\rho}f^D_{\gamma}\\
&\times \oint_{\mathbb{T}^N} \ \Delta_{A_{N-1}} s_{\lambda}(\mathbf{x}) s_{\mu}(\mathbf{x})s_{\rho}(\mathbf{x})s_{\lambda}(\mathbf{x}^{-1})s_{\nu}(\mathbf{x}^{-1})s_{\gamma}(\mathbf{x}^{-1})\prod_{j=1}^N\frac{dx_j}{2\pi ix_j},
\end{aligned}\\[2ex]
\label{sc$SO(2N+1)$21}
I_{SO(2N+1)}^{s=v=0,q=t} (u)  &= 
\begin{aligned}[t]
&\mathcal{X}_N \sum_{\rho\subset S_N^B}\sum_{\gamma\subset S_N^B}\sum_{\lambda}\sum_{\mu,\nu}u^{|\lambda|+\frac{|\mu|+|\nu|+odd(\mu')+odd(\nu')}{2}}f^B_{\rho}f^B_{\gamma}\\
&\times \oint_{\mathbb{T}^N} \ \Delta_{A_{N-1}} s_{\lambda}(\mathbf{x}) s_{\mu}(\mathbf{x})s_{\rho}(\mathbf{x})s_{\lambda}(\mathbf{x}^{-1})s_{\nu}(\mathbf{x}^{-1})s_{\gamma}(\mathbf{x}^{-1})\prod_{j=1}^N\frac{dx_j}{2\pi ix_j}.
\end{aligned}
\end{align}
According to the normalization of the Schur function:
\begin{align}
\label{scproofSchur}
\frac{1}{N!} \oint_{\mathbb{T}^N} \prod_{i=1}^N \frac{dx_i}{2\pi i x_i} 
\Delta_{A_{N-1}}(\mathbf{x}) s_\lambda(\mathbf{x}) s_\mu(\mathbf{x}^{-1}) 
= \delta_{\lambda\mu},
\end{align}
we can get total combinatorial formulas for these indices:
\begin{align}
I_{Sp(2N)}^{p=v=0,q=t} (u)  &=
\begin{aligned}[t]
    &N!\mathcal{X}_N \sum_{\rho\subset S_N^C}\sum_{\gamma\subset S_N^C}\sum_{\lambda,\kappa,\theta,\tau}\ \ \sum_{\text{$\mu$,$\nu$ is even}}u^{|\lambda|+\frac{|\mu|+|\nu|}{2}}f^C_{\rho}f^C_{\gamma}c_{\lambda\mu}^{\kappa}c_{\kappa\rho}^{\tau}c_{\lambda\nu}^{\theta}c_{\theta\gamma}^{\tau},
\label{sc$Sp(2N)$3}
\end{aligned}\\[2ex]
\label{sc$SO(2N+1)'$3}
I_{SO(2N+1)}^{'s=v=0,q=t} (u)  &= 
\begin{aligned}[t]
&N!\mathcal{X}_N \sum_{\rho\subset S_N^B}\sum_{\gamma\subset S_N^B}\sum_{\lambda,\kappa,\theta,\tau}\ \ \sum_{\mu,\nu}u^{|\lambda|+\frac{|\mu|+|\nu|}{2}}f^B_{\rho}f^B_{\gamma}c_{\lambda\mu}^{\kappa}c_{\kappa\rho}^{\tau}c_{\lambda\nu}^{\theta}c_{\theta\gamma}^{\tau},
\end{aligned}\\[2ex]
\label{sc$SO(2N)$3}
I_{SO(2N)}^{s=v=0,q=t} (u)  &=
\begin{aligned}[t]
&N!\mathcal{X}_N \sum_{\rho\subset S_N^D}\sum_{\gamma\subset S_N^D}\sum_{\lambda,\kappa,\theta,\tau}\ \ \sum_{\substack{\text{$m_i(\mu),m_i(\nu)$} \\[3pt] \text{is even}}}u^{|\lambda|+\frac{|\mu|+|\nu|}{2}}f^D_{\rho}f^D_{\gamma}c_{\lambda\mu}^{\kappa}c_{\kappa\rho}^{\tau}c_{\lambda\nu}^{\theta}c_{\theta\gamma}^{\tau},
\end{aligned}\\[2ex]
\label{sc$SO(2N+1)$23}
I_{SO(2N+1)}^{s=v=0,q=t} (u)  &= 
\begin{aligned}[t]
&N!\mathcal{X}_N \sum_{\rho\subset S_N^B}\sum_{\gamma\subset S_N^B}\sum_{\lambda,\kappa,\theta,\tau}\ \ \sum_{\mu,\nu}u^{|\lambda|+\frac{|\mu|+|\nu|+odd(\mu')+odd(\nu')}{2}}f^B_{\rho}f^B_{\gamma}c_{\lambda\mu}^{\kappa}c_{\kappa\rho}^{\tau}c_{\lambda\nu}^{\theta}c_{\theta\gamma}^{\tau}.
\end{aligned}
\end{align}
It would be interesting to establish the equality between (\ref{sc$Sp(2N)$3}) and (\ref{sc$SO(2N+1)$23}) motivated by S-duality in physics. The finiteness of $S_N^B$, $S_N^C$, and $S_N^D$ for finite $N$ enables S-duality verification at low orders in $u$ for small $N$:
\begin{align*}
    I_{Sp(2)}^{p=v=0,q=t}&=\frac{1}{2}\left(2+2u^2+2u^4+2u^6+\mathcal{O}\left(u^7\right)\right), \\
    I_{Sp(4)}^{p=v=0,q=t}&= \frac{1}{4}\left(4+4u^2+8u^4+\mathcal{O}\left(u^5\right)\right),\\
    I_{Sp(6)}^{p=v=0,q=t}&=\frac{1}{8}\left(8+8u^2+0u^3+\mathcal{O}\left(u^4\right)\right), \\
      I_{SO(3)}^{p=v=0,q=t}&=\frac{1}{2}\left(2+2u^2+2u^4+2u^6+\mathcal{O}\left(u^7\right)\right), \\
    I_{SO(5)}^{p=v=0,q=t}&= \frac{1}{4}\left(4+4u^2+8u^4+\mathcal{O}\left(u^5\right)\right),\\
    I_{SO(7)}^{p=v=0,q=t}&=\frac{1}{8}\left(8+8u^2+0u^3+\mathcal{O}\left(u^4\right)\right).
\end{align*}
Furthermore, by exploiting the symmetry under the exchange of parameters $t $ and $ u$, we can immediately obtain analogous results in the $q = u$ limit through the formal substitution $u \to t $ in all relevant expressions. If we start with the original integral representations of deformed Schur indices and take $u=q$, then use the fact \cite{MR2255138}:
\begin{align*}
        \sum_{\substack{\text{$\lambda$ } \\[3pt] \text{$odd(\lambda)$ is even}}}t^{\frac{|\lambda|-odd(\lambda)}{2}}d_{\lambda}(t^{-1})P_{\lambda}(\mathbf{x};t^{-1})&=\prod_{i \geq 0}\frac{1-x_i^2}{1-tx_i^2}\prod_{i<j}\frac{1-x_ix_j}{1-tx_ix_j},\\[10pt]
        \sum_{\lambda}t^{\frac{|\lambda|+odd(\lambda')}{2}}d_{\lambda}(t^{-1})P_{\lambda}(\mathbf{x};t^{-1})&=\prod_{i \geq 0}\frac{1-x_i}{1-tx_i}\prod_{i<j}\frac{1-x_ix_j}{1-tx_ix_j},\\[10pt]
        \sum_{\substack{\lambda \\ \text{$\lambda'$ is even}}}t^{\frac{|\lambda|}{2}}c_{\lambda}(t^{-1})P_{\lambda}(\mathbf{x};t^{-1})&=\prod_{i<j}\frac{1-x_ix_j}{1-tx_ix_j},\\[10pt]
        \sum_{\ell(\lambda)\leq N}t^{|\lambda|}s_{\lambda}(\mathbf{x})s_{\lambda}(\mathbf{x^{-1}})&=\prod_{i,j=1}^N\frac{1}{1-tx_i/x_j},
\end{align*}
where $P_{\lambda}(\mathbf{x};t):=P_{\lambda}(\mathbf{x};0,t)$ is Hall-Littlewood polynomials and
\begin{align*}
    c_{\lambda}(t)&=\prod_{i\geq 1}(1-t)(1-t^3)\cdots(1-t^{m_i(\lambda)-1}),\\
    d_{\lambda}(t)&=\prod_{i\geq 1}(1-t)(1-t^3)\cdots(1-t^{2\left\lceil m_i(\lambda)/2 \right\rceil-1}),
\end{align*}
$\lceil\cdot\rceil$ is the usual ceiling function. By introducing generalized inverse Kostka polynomials $\Bar{K}_{\mu \nu}^{\lambda}(t)$\cite{MR3820363}:
\begin{align*}
   s_{\mu}(\mathbf{x})P_{\nu}(\mathbf{x};t)=\sum_{\lambda}\Bar{K}_{\mu \nu}^{\lambda}(t)s_{\lambda}(\mathbf{x}),  
\end{align*}
then the indices in this limit become:
\begin{align}
I_{Sp(2N)}^{p=v=0,q=u} (t)  &=
\begin{aligned}[t]
    &\frac{1}{2^N}  \sum_{\lambda,\rho}\sum_{\substack{\text{$\mu$, $\nu$} \\[1pt] \text{$odd(\nu)$, $odd(\nu)$} \\[1pt] \text{is even}}}t^{|\lambda|+\frac{|\mu|-odd(\mu)+|\nu|-odd(\nu)}{2}}d_{\mu}(t^{-1})d_{\nu}(t^{-1})\Bar{K}_{\lambda \mu}^{\rho}(t^{-1})\Bar{K}_{\lambda \nu}^{\rho}(t^{-1}),
\label{sc$Sp(2N)$4}
\end{aligned}\\
\label{sc$SO(2N)$4}
I_{SO(2N)}^{s=v=0,q=u} (t)  &=
\begin{aligned}[t]
&\frac{1}{2^{N-1}}\sum_{\lambda,\rho}\sum_{\substack{\mu,\nu \\ \text{$\mu', \nu'$ is even}}}t^{|\lambda|+\frac{|\mu|+|\nu|}{2}}c_{\mu}(t^{-1})c_{\nu}(t^{-1})\Bar{K}_{\lambda \mu}^{\rho}(t^{-1})\Bar{K}_{\lambda \nu}^{\rho}(t^{-1}),\\
\end{aligned}\\[2ex]
\label{sc$SO(2N+1)$24}
I_{SO(2N+1)}^{s=v=0,q=u} (t)  &= 
\begin{aligned}[t]
&\frac{1}{2^N} \sum_{\lambda,\rho}\sum_{\mu,\nu}t^{|\lambda|+\frac{|\mu|+|\nu|+odd(\mu')+odd(\nu')}{2}}d_{\mu}(t^{-1})d_{\nu}(t^{-1})\Bar{K}_{\lambda \mu}^{\rho}(t^{-1})\Bar{K}_{\lambda \nu}^{\rho}(t^{-1}),\\
\end{aligned}
\end{align}
from the symmetry under the exchange of parameters $t$ and $u$, there will be a set of non-trivial identities relating Littlewood-Richardson coefficients to the generalized inverse Kostka polynomials.

For the $q\to0$ limit, Macdonald polynomials reduce to Hall-Littlewood polynomials, For the $BCD_N$ type case, this reduction extends to Koornwinder polynomials as demonstrated in \cite{MR4259867}. In another limit $u=q/t$ the index will reduce to so called the flavor Schur index, further specialization to $t = q^{1/2}$ (i.e., $u = q^{1/2}$), the resulting index is nothing but the original Schur index. There are other methods for calculations in the literature using e,g, modularity. Nevertheless, deriving simplified combinatorial expressions directly from our formulas for these limiting cases is quite challenging.

We can compare our approach with other methods. The character expansion is a powerful method for computing the index \cite{Dolan:2007rq}. For the $U(N)$ case, as discussed in \cite{hatsuda2025deformedschurindicesmacdonald}, the use of Macdonald polynomials has many advantages over the character expansion method. For examples, the character  does not have a general explicit formula and it is hard to find exact expressions of the index. 

For the $BCD$-type gauge groups, the character expansion was discussed in \cite{Sei:2023fjk}, and was used in \cite{du2023schurindicesmathcaln4superyangmills} to compute the simpler unflavored Schur index. For the current calculations of deformed Schur index, our results are more complicated and the advantages over the character expansion are not so clear. Nevertheless, inspired by the $U(N)$ case in \cite{hatsuda2025deformedschurindicesmacdonald}, we believe the current approach using Koornwinder polynomials and Macdonald polynomials are more promising for obtaining exact expressions which would be useful for many purposes such as checking the giant graviton expansions. 

Another powerful method for computing the index is the fermion gas method. In particular, it is useful for obtaining the exact unflavored Schur index \cite{Bourdier:2015wda}. However this method requires the use of the Cauchy determinant formula for Jacobi theta functions, which seems not available in the current context of deformed Schur index. 

We can provide some more details for the character expansion as the followings
\begin{align*}
    &I_{\mathrm{USp} (2N)} (x) = \sum_{\lambda} \left[ \frac{ f_{\lambda}(x) }{z_{\lambda}} \frac{1}{2^{l(\lambda)}} \sum_{\tilde{\lambda} \in \mathrm{Ev} (\lambda)} \sum_{\mu \in R_{2N}^{\,\mathrm{c}} (2|\lambda|)} \chi_{\mu}^S (\tilde{\lambda}) \right],\\
    &I_{\mathrm{SO} (n)} (x) = \sum_{\lambda} \left[ \frac{ f_{\lambda}(x) }{z_{\lambda}} \frac{1}{2^{l(\lambda)}} \sum_{\tilde{\lambda} \in \mathrm{Ev} (\lambda)} (-1)^{l(\tilde\lambda)}\sum_{\mu \in R_{n}^{\,\mathrm{r}} (2|\lambda|) \cup W_{n}^{\,\mathrm{r}} (2 \lvert\lambda \rvert )} \chi_{\mu}^S (\tilde{\lambda}) \right],\\
    &I_{\mathrm{SO} (2N + 1)} (x) = \sum_{\lambda} \left[ \frac{ f_{\lambda}(x) }{z_{\lambda}} \frac{1}{2^{l(\lambda)}} \sum_{\tilde{\lambda} \in \mathrm{Ev} (\lambda)} (-1)^{l(\tilde\lambda)}\sum_{\mu \in R_{2N + 1}^{\,\mathrm{r}} (2|\lambda|)} \chi_{\mu}^S (\tilde{\lambda}) \right].
\end{align*}
$x$ here denotes the set of all fugacities $\{t,u,v,p,q\}$, and
\begin{align*}
f_n(t,u,v,p,q)&=\frac{t^n+u^n-t^n u^n-q^n}{1-q^n},\\
f_{\lambda}(t,u,v,p,q)&=f_{\lambda_1}f_{\lambda_2}\cdots f_{\lambda_{l(\lambda)}}(t,u,v,p,q).
\end{align*}
The notations for the Young tableau are explained in  \cite{Sei:2023fjk, du2023schurindicesmathcaln4superyangmills}.

Compared to the character expansion method, our integral representation yields more compact formulas in the $u=0$ limit. Furthermore, in the $q=t$ and $q=u$ limits, the function $f_n(t, u, v, p, q)$ reduces to $u^n$ and $t^n$, respectively. The corresponding summations over subsets of Young tableaux in the character expansion are similar to the current results. The indices in $q=t$ and $q=u$  limits can be obtained by interchanging the parameters $t$ and $u$ in the character expansion method, which is not manifest in our case.

\section{Conclusion}
\label{sec6}

In this paper, we study the deformed Schur indices of $\mathcal{N}=4$ SYM theories with $BCD$-type gauge groups. We develop a computational method based on symmetric functions and introduce a novel integral formulation involving Koornwinder weight functions and Koornwinder polynomials.  We compute the integrals in various expansion schemes and  further examine several limiting cases where the results are much simplified.

Although we have made some progress in the calculations, the results are much complicated than the $SU(N)$ case considered in \cite{hatsuda2025deformedschurindicesmacdonald}. It is interesting to try to further simplify the formulas. Some close expressions would be much helpful in testing giant graviton expansions. It would be interesting to find connections with other well known combinatorial problems, such as the generalized MacMahonÕs sum-of-divisors functions as appearing in the calculations of unflavored Schur index \cite{Kang:2021lic}. Some direct mathematical proofs of the non-trivial identities resulting from S-duality or the symmetry of the fugacity parameters may help to elucidate further calculations. 

It would be interesting to apply the techniques to the calculations of the most general 1/16 BPS superconformal indices of $\mathcal{N}=4$ SYM theories, which has one more fugacity parameter and is relevant for counting black hole microstates in AdS$_5$. Such results would provide more precise counting than previous works, help to elucidate the connections between the different approaches in the works \cite{Cabo-Bizet:2018ehj, Choi:2018hmj, Benini:2018ywd}. 

It would be also interesting to compute deformed Schur indices of $BCD$-type gauge groups with defects. The giant graviton expansion with defects has been studied recently in e.g. \cite{Imamura:2024lkw, Beccaria:2024oif, Beccaria:2024dxi, Imamura:2024zvw}. The Schur indices with defects are very useful to understand the characters of the corresponding vertex operator algebras and their modularity, see e.g. in \cite{Zheng:2022zkm, Pan:2024bne, Pan:2024hcz}. We hope to further study these issues in the future.

\section*{Acknowledgements}
We thank Xin Wang for helpful discussions. We are supported by the National Natural Science Foundation of China Grants No.12325502 and No.12247103.

\appendix
\section{Young diagram and symmetric function}
\label{appendixA}
\subsection{Young diagram and notations}
This chapter provides formal definitions for the notations referenced without explicit specification in the main text.
Some notation in young diagram: Given a partition $\lambda = (\lambda_1, \lambda_2, . . .)$, the young diagram is ,for example:

\begin{table}[h]
\centering
\begin{tabular}{lllllllll}
\cline{1-8}
\multicolumn{1}{|l|}{}                         & \multicolumn{1}{l|}{}                         & \multicolumn{1}{l|}{}                         & \multicolumn{1}{l|}{\cellcolor[HTML]{6665CD}} & \multicolumn{1}{l|}{}                         & \multicolumn{1}{l|}{}                         & \multicolumn{1}{l|}{}                         & \multicolumn{1}{l|}{} &  \\ \cline{1-8}
\multicolumn{1}{|l|}{\cellcolor[HTML]{00D2CB}} & \multicolumn{1}{l|}{\cellcolor[HTML]{00D2CB}} & \multicolumn{1}{l|}{\cellcolor[HTML]{00D2CB}} & \multicolumn{1}{l|}{\cellcolor[HTML]{FFFE65}} & \multicolumn{1}{l|}{\cellcolor[HTML]{FE0000}} & \multicolumn{1}{l|}{\cellcolor[HTML]{FE0000}} & \multicolumn{1}{l|}{\cellcolor[HTML]{FE0000}} &                       &  \\ \cline{1-7}
\multicolumn{1}{|l|}{}                         & \multicolumn{1}{l|}{}                         & \multicolumn{1}{l|}{}                         & \multicolumn{1}{l|}{\cellcolor[HTML]{9AFF99}} & \multicolumn{1}{l|}{}                         & \multicolumn{1}{l|}{}                         & \multicolumn{1}{l|}{}                         &                       &  \\ \cline{1-7}
\multicolumn{1}{|l|}{}                         & \multicolumn{1}{l|}{}                         & \multicolumn{1}{l|}{}                         & \multicolumn{1}{l|}{\cellcolor[HTML]{9AFF99}} & \multicolumn{1}{l|}{}                         &                                               &                                               &                       &  \\ \cline{1-5}
\multicolumn{1}{|l|}{}                         & \multicolumn{1}{l|}{}                         & \multicolumn{1}{l|}{}                         & \multicolumn{1}{l|}{\cellcolor[HTML]{9AFF99}} &                                               &                                               &                                               &                       &  \\ \cline{1-4}
\multicolumn{1}{|l|}{}                         & \multicolumn{1}{l|}{}                         & \multicolumn{1}{l|}{}                         &                                               &                                               &                                               &                                               &                       &  \\ \cline{1-3}
\multicolumn{1}{|l|}{}                         &                                               &                                               &                                               &                                               &                                               &                                               &                       &  \\ \cline{1-1}
                                               &                                               &                                               &                                               &                                               &                                               &                                               &                       &  \\
                                               &                                               &                                               &                                               &                                               &                                               &                                               &                       & 
\end{tabular}
\caption{Young diagram for partition [8,7,7,5,4,3,1]}\label{table1}
\end{table}
The coordinate of box is denoted by $(i,j)$, the coordinate of yellow box in Table \ref{table1} is $s=(4,2)$, The arm-length $a(s)$, leg-length $l(s)$, arm-colength $a'(s)$ and leg-colength $l'(s)$ of the square $s = (4,2) \in \lambda$ are defined by the number of the red, green, blue, purple boxes:
\begin{align}
\begin{split}
&a(s) = a_\lambda(s) := \lambda_i - j,\quad\quad  a'(s) = a'_\lambda(s) := j - 1, \\
&l(s) = l_\lambda(s) := \lambda_j' - i,\quad\quad\ \  l'(s) = l'_\lambda(s) := i - 1,
\end{split}
\end{align}
where $\lambda'=(\lambda_1', \lambda_2', . . .)$ is the transpose young diagram of $\lambda$. 

\subsection{Symmetric function}
We only introduce monomial symmetric functions\cite{noumi2023macdonald} in this subsection. Consider an arbitrary polynomial $f(\bm{x}) \in \mathbb{C}[x_1,\ldots,x_n]$ in the variables $\bm{x} = (x_1,\dots,x_n)$, expressed as a finite sum:
\begin{equation*}
f(\bm{x}) = \sum_{\mu_1,\dots,\mu_n \geq 0} a_{\mu_1,\dots,\mu_n} x_1^{\mu_1} \cdots x_n^{\mu_n}.
\end{equation*}
The action of a permutation $\sigma \in \mathbf{S}_n$ on $f$ is defined by
\begin{equation*}
\sigma(f) = f(x_{\sigma(1)},\ldots,x_{\sigma(n)}) = \sum_{\mu_1,\dots,\mu_n \geq 0} a_{\mu_1,\dots,\mu_n} x_{\sigma(1)}^{\mu_1} \cdots x_{\sigma(n)}^{\mu_n}.
\end{equation*}
We employ the notation
\begin{equation}
\bm{x}^{\bm{\mu}} = x_1^{\mu_1} \cdots x_n^{\mu_n}, \quad\quad\quad \deg_{\bm{x}} \bm{x}^{\bm{\mu}} = |\bm{\mu}| = \mu_1 + \cdots + \mu_n,
\end{equation}
and note that the action of $\sigma \in \mathbf{S}_n$ on $\bm{x}^{\bm{\mu}}$ is given by
\begin{align*}
\sigma(\bm{x}^{\bm{\mu}})
&= \bm{x}^{\sigma \cdot \bm{\mu}} = x_{\sigma(1)}^{\mu_1} \cdots x_{\sigma(n)}^{\mu_n} = x_1^{\mu_{\sigma^{-1}(1)}} \cdots x_n^{\mu_{\sigma^{-1}(n)}}.
\end{align*}

For each partition $\lambda$, we define
\begin{equation}
m_\lambda(\bm{x}) = \sum_{\mu \in \mathbf{S}_n\cdot \lambda} \bm{x}^\mu,
\end{equation}
as the sum of all monomials corresponding to elements in the orbit $\mathbf{S}_n\cdot \lambda$. This $m_{\lambda(\bm{x})}$ is the monomial symmetric function associated with partition $\lambda$, where each distinct monomial obtained from $\bm{x}^{\lambda}$ by permutation appears exactly once with coefficient 1. Equivalently,
\begin{equation}
m_\lambda(\bm{x}) = \frac{1}{|\mathbf{S}_{n,\lambda}|} \sum_{\sigma \in \mathbf{S}_n} \sigma \cdot \bm{x}^{\lambda} = \frac{1}{|\mathbf{S}_{n,\lambda}|} \sum_{\sigma \in \mathbf{S}_n} \bm{x}^{\sigma \cdot \lambda},
\end{equation}
where $\mathbf{S}_{n,\lambda} = \{\sigma \in \mathbf{S}_n \mid \sigma\cdot\lambda = \lambda\}$ denotes the stabilizer subgroup of $\lambda$.

\section{Macdonald polynomial expansion of symmetric function products}
\label{appendixB}
In this section, we provide some results of the Macdonald polynomial expansion of symmetric function products used in calculation of the first few orders:
\begin{flalign*}
&m_{[1]}  = P_{[1]},\quad\quad m_{[1]}  P_{[1]} = \left( \frac{qt+q-t-1}{qt-1} \right) P_{[1,1]} + P_{[2]},&\\[1ex]
&m_{[1]}P_{[1,1]}  = \scalebox{1.3}{$\frac{-qt^2-qt+t^2-q+t+1}{-qt^2+1}$}P_{[1,1,1]}+P_{[2,1]},\ m_{[1]}  P_{[2]} = \frac{-q^3t^2+qt^2+q^2-1}{-q^3t^2+q^2t+qt-1}  P_{[2,1]} + P_{[3]},&\\[1ex]
&m_{[1]}  P_{[1,1,1]} = \left( \frac{qt^3+qt^2-t^3+qt-t^2+q-t-1}{qt^3-1} \right) P_{[1,1,1,1]} + P_{[2,1,1]},& \\[1ex]
&m_{[1]}  P_{[2,1]} =\scalebox{1.2}{$\frac{-q^4t^5-q^4t^4+q^3t^5+q^3t^4+q^2t^4+q^3t^2+q^2t^3-qt^4+q^3t-q^2t^2-qt^3-q^2t-qt-q+t+1}{-q^4t^5+q^3t^4+q^3t^3+q^2t^3-q^2t^2-qt^2-qt+1}$}   P_{[2,1,1]} &\\[1ex]
&\quad\quad\quad\quad+  \frac{qt+q-t-1}{qt-1} P_{[2,2]} + P_{[3,1]},& \\[1ex]
&m_{[1]}  P_{[3]} = \left( \frac{-q^5t^2+q^2t^2+q^3-1}{-q^5t^2+q^3t+q^2t-1} \right) P_{[3,1]} + P_{[4]},& \\[1ex]
&m_{[1,1]}  = P_{[1,1]},\quad\quad m_{[1,1]}  P_{[1]} = \left( \frac{-q t^2 - q t + t^2 - q + t + 1}{-q t^2 + 1} \right) P_{[1,1,1]} + P_{[2,1]},& \\[1ex]
& m_{[1,1]}  P_{[1,1]} = \scalebox{1.2}{$\frac{q^2 t^5 + q^2 t^4 - q t^5 + 2 q^2 t^3 - 2 q t^4 + q^2 t^2 - 3 q t^3 + t^4 + q^2 t - 3 q t^2 + t^3 - 2 q t + 2 t^2 - q + t + 1}{q^2 t^5 - q t^3 - q t^2 + 1}$} P_{[1,1,1,1]} &  \\[1ex]
& \quad\quad\quad\quad\  + \left( \frac{q t + q - t - 1}{q t - 1} \right) P_{[2,1,1]} + P_{[2,2]},& \\[1ex]
&m_{[1,1]} \cdot P_{[2]} = \left( \frac{q^3 t^3 - q t^3 - q^2 + 1}{q^3 t^3 - q^2 t^2 - q t + 1} \right) P_{[2,1,1]} + P_{[3,1]},\ m_{[2]}=\left( \frac{-q \cdot t + q - t + 1}{q \cdot t - 1} \right) P_{[1,1]} + P_{[2]},& \\[1ex]
&m_{[2]}P_{[1]}=\left( \frac{q^2 t^3 - t^3 - q^2 + 1}{-q^2 t^3 + q t^2 + q t - 1} \right) P_{[1,1,1]} 
  + \left( \frac{q^2 - q t + q - t}{q^2 t - 1} \right) P_{[2,1]} 
  + P_{[3]},& \\[1ex]
&m_{[2]}P_{[1,1]}=\frac{-q^2 t^5 - q^2 t^3 + t^5 + q^2 t^2 + t^3+ q^2 - t^2 - 1}
  {q^2 t^5 - q t^3 - q t^2 + 1} P_{[1,1,1,1]} 
  + \frac{q^3 t - q^2 t^2 - q t + t^2}{q^3 t^3 - q^2 t^2 - q t + 1}  P_{[2,1,1]}&\\[1ex]
  &\quad\quad\quad\quad + \left( \frac{-q t + q - t + 1}{q t - 1} \right) P_{[2,2]} 
  + P_{[3,1]},&\\[1ex]
  \\
&m_{[2]}P_{[2]}= \scalebox{1.3}{$\frac{-q^4 t^4 + q^4 t^3 - q^3 t^4 + q^3 t^3 + q^2 t^4- q^2 t^3 + q t^4 + q^3 t - q t^3 - q^3 + q^2 t - q^2 - q t + q - t + 1}{q^4 t^4 - 2 q^3 t^3 + 2 q t - 1}$}  P_{[2,1,1]} & \\[1ex]
&\quad\quad\quad +\scalebox{1.3}{$\frac{q^4 t^3 + q^4 t^2 - q^3 t^3 - q^3 t^2 - q^2 t^3- q^3 t - q^2 t^2 + q t^3 - q^3 + q^2 t + q t^2 + q^2 + q t + q - t - 1}{q^4 t^3 - 2 q^3 t^2 - q^2 t^2 + q^2 t + 2 q t - 1}$}   P_{[2,2]} & \\[1ex]
&\quad\quad\quad + \frac{q^3 t - q^2 t^2 - q^3 + 2 q^2 t - q t^2 - q^2 + q t}{q^4 t^2 - q^3 t - q t + 1}  P_{[3,1]} 
  + P_{[4]},&\\[1ex]
&m_{[1,1,1]}  =P_{[1,1,1]},\quad\quad m_{[1,1,1]}  P_{[1]} =  \frac{qt^3+qt^2-t^3+qt-t^2+q-t-1}{qt^3-1}  P_{[1,1,1,1]} + P_{[2,1,1]},& \\[1ex]
&m_{[2,1]}  =\frac{2qt^2-qt+t^2-q+t-2}{-qt^2+1}P_{[1,1,1]}+ P_{[2,1]},&\\[1ex]
&m_{[2,1]}  P_{[1]} = \frac{-2q^2t^5-q^2t^4+qt^5-qt^4+t^5+2t^4+2q^2t+q^2-qt+q-t-2}{q^2t^5-qt^3-qt^2+1}  P_{[1,1,1,1]}&\\[1ex]
&\quad\quad\quad\quad+\frac{-q^3t^3+2q^3t^2-2q^2t^3+q^3t+t^2-2q+2t-1}{q^3t^3-q^2t^2-qt+1}P_{[2,1,1]}+\frac{qt+q-t-1}{qt-1}P_{[2,2]}+P_{[3,1]},& \\[1ex]
&m_{[3]}  = \frac{-q^2t^3+q^2t^2-qt^3+q^2t+qt^2-t^3-q^2+qt+t^2-q+t-1}{-q^2t^3+qt^2+qt-1}P_{[1,1,1]}&\\[1ex]
&\quad\quad+\frac{q^2t-q^2+qt-q+t-1}{-q^2t+1}  P_{[2,1]}+P_{[3]},&\\[1ex]
&m_{[3]}  P_{[1]} = \frac{q^3t^6-q^3t^4-t^6-q^3t^2+t^4+q^3+t^2-1}{q^3t^6-q^2t^5-q^2t^4-q^2t^3+qt^3+qt^2+qt-1}P_{[1,1,1,1]}&\\[1ex]
&\quad\quad\quad+ \frac{-q^3t^2+q^2t^3-q^2t^2+qt^3+q^3-q^2t-qt^2+t^3+q^2-qt+q-t}{q^3t^3-q^2t^2-qt+1}P_{[2,1,1]}&\\[1ex]
&\quad\quad\quad+\frac{-q^3t^2+q^3+t^2-1}{q^3t^2-q^2t-qt+1}P_{[2,2]}+\frac{q^3-q^2t+q^2-qt+q-t}{q^3t-1}P_{[3,1]}+P_{[4]},& \\[1ex]
& m_{[2,1,1]}  =  \left( \frac{-3 q t^3 + q t^2 - t^3 + q t - t^2 + q - t + 3}{q t^3 - 1} \right) P_{[1,1,1,1]} + P_{[2,1,1]},&\\[1ex]
& m_{[2,2]}   =  \left( \frac{q^2 t^5 - q^2 t^4 + q t^5 - q^2 t^2 - q t^3 + t^4 + q^2 t - q t^2 - t^3 + q - t + 1}{q^2 t^5 - q t^3 - q t^2 + 1} \right) P_{[1,1,1,1]} &\\[1ex]
& \quad\quad\ \ + \left( \frac{-q t + q - t + 1}{q t - 1} \right) P_{[2,1,1]} + P_{[2,2]}.&\\
\end{flalign*}
\section{Derivation of (\ref{KDSI})}\label{app3}
In this section, we establish the equality between equations (\ref{Cdsi}), (\ref{Ddsi}) and (\ref{KDSI}) under parameter specialization. We focus on the $SO(2N)$ case, noting that other cases follow analogously.

In specialization:
\begin{align*}
    (t_0,t_1,t_2,t_3)= (-1,+1,-\sqrt{q},+\sqrt{q}).
\end{align*}
To establish the equivalence, the required identities are:
\begin{align*}
\label{qsym}
    &(\sqrt{q}a;q)_\infty  (-\sqrt{q}a;q)_\infty=(qa^2;q^2)_\infty,\ (a;q)_\infty  (-a;q)_\infty=(a^2;q^2)_\infty,\ (qa;q^2)_\infty(a;q^2)_\infty=(a;q)_\infty,
\end{align*}
the weight function now is:
\begin{align*}
    \prod_{1 \leq i \leq n} \frac{(x_i^{2};q)_\infty}
{(+ x_i, - x_i, \sqrt{q} x_i, -\sqrt{q} x_i;q)_\infty} = 1,
\end{align*}
so
\begin{align*}
   \Delta_n(x_1,x_2,\ldots,x_n;q,t;1,-1,+\sqrt{q},-\sqrt{q}) &=\prod_{1 \leq i < j \leq n} \frac{(x_i x_j;q)_\infty}
{(t x_i x_j^;q)_\infty}\prod_{1 \leq i < j \leq n} \frac{(x_i x_j^{-1};q)_\infty}
{(t x_i x_j^{-1};q)_\infty},\\
\Delta_n(x_1^{-1},x_2^{-1},\ldots,x_n^{-1};q,t;1,-1,+\sqrt{q},-\sqrt{q}) &=\prod_{1 \leq i < j \leq n} \frac{(x_i^{-1} x_j^{-1};q)_\infty}
{(t x_i^{-1} x_j^{-1};q)_\infty}\prod_{1 \leq i < j \leq n} \frac{(x_i^{-1} x_j;q)_\infty}
{(t x_i^{-1} x_j;q)_\infty}.
\end{align*}
‌Multiply them together
\begin{equation}\label{aweight}
     \Delta_n(\mathbf{x};q,t;1,-1,+\sqrt{q},-\sqrt{q}) \Delta_n(\mathbf{x^{-1}};q,t;1,-1,+\sqrt{q},-\sqrt{q})=\prod_{1 \leq i < j \leq N} \frac{(x_i^{\pm 1} x_j^{\pm 1}; q)_{\infty}}{(t x_i^{\pm 1} x_j^{\pm 1}; q)_{\infty}}. 
\end{equation}
Consider Theorem.\ref{thm1} with $\lambda=\emptyset$, due to $P_{\lambda}(\mathbf{x};q,t)=1$, then change variables $\mathbf{x} \to \sqrt{u}\mathbf{x}$, $\mathbf{x}^{-1} \to \sqrt{u}\mathbf{x}^{-1} $, it becomes:
\begin{align*}
&\lim_{m \to \infty} u^{\frac{mn}{2}}(x_1 \cdots x_n)^m K_{m^n}(\sqrt{u}\mathbf{x}; q, t; 1,-1, \sqrt{q}, -\sqrt{q})
=\frac{(t u x_i x_j; q)_{\infty}}{( u x_i x_j; q)_{\infty}},\\
&\lim_{m \to \infty} u^{\frac{mn}{2}}(x_1 \cdots x_n)^{-m} K_{m^n}(\sqrt{u}\mathbf{x}^{-1}; q, t; 1,-1, \sqrt{q}, -\sqrt{q})
=\frac{(t ux_i^{-1} x_j^{-1}; q)_{\infty}}{( u x_i^{-1} x_j^{-1}; q)_{\infty}},
\end{align*}
combine with
\begin{equation}
\label{macintegrand1}
  \prod_{i,j=1}^N \frac{(tu x_i/x_j;q)_\infty}{(u x_i/x_j;q)_\infty} 
= \sum_{\ell(\lambda) \leq N} u^{|\lambda|} b_\lambda P_\lambda(\mathbf{x};q,t) P_\lambda(\mathbf{x}^{-1};q,t)=\sum_{\ell(\lambda) \leq N} b_\lambda P_\lambda(\sqrt{u}\mathbf{x};q,t) P_\lambda(\sqrt{u}\mathbf{x}^{-1};q,t),  
\end{equation}
change $x\to z$, we get
\begin{align}\label{main}
\begin{split}
     \lim_{m \to \infty} &\sum_{\ell(\lambda) \leq N}(z_1 \cdots z_n)^m K_{m^n}(\sqrt{u}\mathbf{z}; q, t;  \pm1, \pm\sqrt{q})(z_1 \cdots z_n)^{-m} K_{m^n}(\sqrt{u}\mathbf{z}^{-1}; q, t; \pm1, \pm\sqrt{q})\\
    &\quad\quad\quad\quad\times u^{mn}b_\lambda P_\lambda(\sqrt{u}\mathbf{z};q,t) P_\lambda(\sqrt{u}\mathbf{z}^{-1};q,t)=\prod_{1 \leq i < j \leq N}\frac{( t u z_i^{\pm 1} z_j^{\pm 1}; q)_{\infty}}{(u z_i^{\pm 1} z_j^{\pm 1}; q)_{\infty}}\times \frac{(tu;q)^N_\infty}{(u;q)^N
    _\infty},
\end{split}
\end{align}
so (\ref{aweight}) times (\ref{main}) is same as the integrand in (\ref{Ddsi}) up to an overall factors. Combine with Theorem.\ref{thm2}, we can get (\ref{KoornwidnerSO(2N)aftertheorem2}). The $C_N$ case is similar, but in $B_N$ case, we can't find the same formula like (\ref{KDSI}), but using Theorem.\ref{thm2}, we can get the integral (\ref{KoornwidnerSO(2N+1)aftertheorem2}) which is similar to other case.




\bibliographystyle{JHEP} 
\bibliography{References} 
\end{document}